\documentclass[preprint,eqsecnum,showpacs,nofootinbib,aps,amsmath,amssymb%
              ,tightenlines]{revtex4}
\usepackage{amsmath,amssymb,amsbsy,amsthm,latexsym,graphics}
\usepackage{vmargin} 
\usepackage{enumerate}
\usepackage{calc}    

\usepackage{colordvi}

\newtheorem{theorem}{Theorem}[section] 

\newtheorem{lem}[theorem]{Lemma}
\newtheorem{cor}[theorem]{Corollary}
\newtheorem{defn}[theorem]{Definition}
\newtheorem{cond}[theorem]{Condition}   

\newcounter{mnotecount}[section]

\newcommand{\M}{\mathcal{M}} 
\newcommand{\ih}{\triangle}  
\newcommand{\slc}{\tilde{\ih}} 
\newcommand{\bas}{\hat{\ih}}   
\newcommand{\I}{\mathbb{R}} 
\newcommand{\bsl}{\boldsymbol{\ell}} 

\newcommand{\tq}{\tilde{q}} 
\newcommand{\hq}{\hat{q}} 
\newcommand{\Ric}{\mathcal{R}} 
\newcommand{\tD}{\tilde{D}} 
\newcommand{\hD}{\hat{D}} 
\newcommand{\tS}{\tilde{S}}
\newcommand{\hS}{\hat{S}}
\newcommand{\wI}{\omega}
\newcommand{\hwI}{\hat{\wI}}
\newcommand{\twI}{\tilde{\wI}}
\newcommand{\w}[1]{{\wI^{\scriptscriptstyle{(#1)}}}{}}
\newcommand{\tw}[1]{{\twI^{\scriptscriptstyle{(#1)}}}{}}
\newcommand{\hw}[1]{{\hwI^{\scriptscriptstyle{(#1)}}}{}}
\newcommand{\sgr}[1]{\kappa^{(#1)}}

\newcommand{\rd}{{\rm d}}

\newcommand{\lie}{\mathcal{L}}  

\DeclareMathOperator*{\const}{const}

\newcommand{\tRic}{\tilde{\Ric}}
\newcommand{\hRic}{\hat{\Ric}}

\newcommand{\fracs}[2]{{\scriptstyle\frac{#1}{#2}}}
\newcommand{\RicScript}[1]{{}^{{}^{{}^{\scriptscriptstyle{(#1)}}}}}
\newcommand{\Ricn}[1]{\RicScript{#1}\!\!\!\Ric}
\newcommand{\tRicn}[1]{\RicScript{#1}\!\!\!\tRic}
\newcommand{\hRicn}[1]{\RicScript{#1}\!\!\!\hRic}

\newcommand{\Ein}[1]{\RicScript{#1}\!\!\! G}
\newcommand{\Riem}[1]{\RicScript{#1}\!\!\! R}
\newcommand{\bRiem}[1]{\RicScript{#1}\!\!\! \bar{R}}
\newcommand{\Weyl}[1]{\RicScript{#1}\!\!\! C}
\newcommand{\Ricl}{{\Ric^{\scriptscriptstyle{(\ell)}}}{}}
\newcommand{\bRicl}{{\bar{\Ric}^{\scriptscriptstyle{(\ell)}}}{}}
\newcommand{\Bv}[1]{{B^{\scriptscriptstyle{(#1)}}}{}}
\newcommand{\Bvel}{\Bv{\ell}{}}
\newcommand{\Bvlo}{\Bv{\ell_o}{}}

\newcommand{\bq}{\bar{q}}
\newcommand{\bBv}[1]{{\bar{B}^{\scriptscriptstyle{(#1)}}}{}}
\newcommand{\bBvel}{\bBv{\ell}{}}
\newcommand{\bBvlo}{\bBv{\ell_o}{}}
\newcommand{\thv}[1]{{\theta^{\scriptscriptstyle{(#1)}}}{}}
\newcommand{\sgv}[1]{{\sigma^{\scriptscriptstyle{(#1)}}}{}}

\newcommand{\comment}[1]{}

\begin{document}

\title{Quasi-local rotating black holes in higher dimension: geometry}

\author{Jerzy Lewandowski${}^{1,2,3}$}
  \email{lewand@fuw.edu.pl}
\author{Tomasz Pawlowski${}^{1,2,3}$}
  \email{tpawlow@fuw.edu.pl}

\affiliation{
  ${}^1$Instytut Fizyki Teoretycznej\\
    Uniwersytet Warszawski, ul. Hoza 69, 00-681, Warsaw, Poland\\
  ${}^2$Center for Gravitational Physics and Geometry,\\
    Physics Department, Penn State, University Park, PA 16802, USA\\
  ${}^3$Max-Planck-Institut f¨ur Gravitationsphysik,
    Albert-Einstein-Institut,
    Am M¨uhlenberg 1, D-14476 Potsdam, Germany
}

\begin{abstract} With a help of a generalized Raychaudhuri equation
non-expanding null surfaces are studied in arbitrarily dimensional
case. The definition and basic properties of non-expanding and
isolated horizons known in the literature in the $4$ and $3$
dimensional cases are generalized. A local description of
horizon's geometry is provided. The Zeroth Law of black hole
thermodynamics is derived. The constraints have a similar
structure to that of the  $4$ dimensional spacetime case. The
geometry of a  vacuum isolated horizon is determined by the
induced metric and the rotation 1-form potential,
local generalizations of the area and the angular momentum
typically used in the stationary black hole solutions case.
\end{abstract}


\maketitle

\section{Introduction}
The theory of non-expanding and isolated horizons in $4$
dimensional space-time  \cite{newman,abf,abdfklw,abl-g,abl-mech}
is a quasi local approach to black hole in equilibrium. A horizon
is a compact, space-like 2-surface expanding at the speed of
light, however, not changing its area element. No symmetry
assumptions are made about a spacetime neighborhood surrounding
the horizon. In fact, generically there is no Killing vector
\cite{l}. The  parameters characterizing stationary black hole
solutions, like the area and the angular momentum, are replaced by
appropriate local fields \cite{abl-g,aepv}. Despite of this
enormous change in the amount of the degrees of freedom, the
Zeroth and the First law of Black Hole Thermodynamics still hold
(see also \cite{wald-killing-horizon-noether}). The goal of the
current and a coming paper \cite{klp} is a generalization of those
results to the higher dimensional case. Whether or not the
generalization would be straightforward  was a priori not known.
In the calculations concerning the $4$ and $3$ dimensional \cite{W1} cases
the Newman-Penrose formalism (and its adaptation to the $3$
dimensions) was used many times, for example in the proof of the
Zeroth Law.

We consider an  $n$ dimensional space-time of the signature
$(-,+...+)$ and arbitrary $n>2$. First, we derive a higher
dimensional Raychaudhuri equation for a null, geodesic flow. This
is  an easy generalization of the derivation one can find in
\cite{wald-book}.

Next, we study non-expanding null surfaces. Our considerations are
local, therefore the results may be applied to the surfaces of
arbitrary topology. Assuming the usual energy inequalities
(classical), we find that  the vanishing of the expansion of a
null surface implies the vanishing of the shear. In the
consequence,  the space-time covariant derivative preserves the
tangent bundle  of each non-expanding null surface, and induces a
covariant derivative therein. The induced degenerate metric tensor
and the induced covariant derivative (partially independent of each other)
constitute the {\it geometry} of a non-expanding null surface. The
geometry is the subject of our study. The induced degenerate
metric tensor can be locally identified with a  metric tensor
defined on the $n-2$ dimensional  space of the tangent null
curves. We do not find any restrictions on that $n-2$ metric
tensor. The rotation of a given non-expanding null surface is
described by a differential 2-form  invariant derived from the
covariant derivative, the  {\it rotation 2-form}.  Its properties
imply the Zeroth Law upon quite week energy conditions. The
remaining components of the surface covariant derivative --briefly
speaking, the shear and expansion of a transversal null vector
field-- are subject to constraint equations which dictate a null
evolution along the surface.

The constraint equations become particularly important in the case
of a surface admitting a null symmetry, called isolated null
surface. Due to them, in the vacuum case, the whole  geometry of a
given non-extremal null isolated surface is locally characterized
by the induced degenerate metric tensor, the rotation 2-form (or
even by their pullbacks to a spacelike $n-2$ dimensional subsurface)
and the value of the cosmological constant. We also derive the
equations constraining the induced metric and the rotation 2-form
in the vacuum extremal isolated null surface case.

In the last section, we apply our local results to the
non-expanding and isolated horizons, which are defined by assuming
the existence of a global, compact space-like cross-section and
the product structure.

Our characterization of the geometry  is used in a coming paper
\cite{klp} to introduce a canonical framework for the isolated
horizons and to derive the First Law in a way analogous to the $4$
and $3$ dimensional cases.

\subsection{Assumptions and notation convention}

We are considering a manifold $\M$ of the dimension  $n>2$ (our
primary interest is in the case $n>4$). $\M$ is equipped with a
(pseudo) metric tensor field $g_{\alpha\beta}$ of the signature
$(-,+...+)$ (one minus and $n-1$ pluses) and the corresponding
Levi-Civita connection $\nabla_\alpha$. The corresponding
Riemann\footnote{We use the following convention:
$[\nabla_\alpha,\nabla_\beta]X^\gamma=R^{\gamma}{}_{\delta\alpha\beta}
X^\delta$}, Weyl, Ricci and Einstein tensors are denoted,
respectively, by $\Riem{n}^\alpha{}_{\beta\gamma\delta}$,
$\Weyl{n}^\alpha{}_{\beta\gamma\delta}$, $\Ricn{n}_{\alpha\beta}$
and $\Ein{n}_{\alpha\beta}$. We often refer to the Einstein
equations, which read
\begin{equation}
  \Ein{n}_{\alpha\beta}\ =\ -\Lambda g_{\alpha\beta} 
  + T_{\alpha\beta} \ ,
\end{equation}
where  $\Lambda$ is a constant called cosmological and
$T_{\alpha\beta}$ is the matter energy-momentum tensor.

The following (abstract) index notation will be used in this
paper:
\begin{enumerate}[ (i)]
  \item Indexes of the spacetime tensors will (and have already been)
        be denoted by lower Greek letters:
        $\alpha,\beta,\gamma,\delta...$.
  \item Tensors defined in $n-1$ dimensional null subspaces
        (tangent to a null surface except Section \ref{sec:ngf})
        will carry indexes denoted by lower Latin letters:
        $a,b,c,d...$.
  \item Capital Latin letters $A,B,C,D,...$  will be used as the
        indexes of tensors considered in $n-2$ dimensional spaces
        (the quotient of a null space by the null direction, the space
        tangent to a spacelike section of a null surface, the space
        tangent to the manifold of null curves in a null surface).
\end{enumerate}

\section{Null geodesic flows, null surfaces}
\label{sec:ngf}

\subsection{Null geodesic flows, generalized Raychaudhuri equation}
\label{sec:ngfR}

Consider a null geodesic vector field $\ell$, that is a null vector
field such that
\begin{equation}\label{eq:geo}
  \nabla_\ell\ell = \sgr{\ell}\ell \ ,
\end{equation}
where $\sgr{\ell}$ is an arbitrary function. We are assuming
that $\ell$ is a section of a  sub-bundle $L$ of the tangent bundle
$T\M$ whose fibers are one dimensional
(the assumption is satisfied by every nowhere-vanishing $\ell$).
It follows from (\ref{eq:geo}), that the sub-bundle
$L^\perp\subset T\M$ consisting of all the vectors tangent to $\M$
and orthogonal to the fibers of $L$ is preserved by the flow of
the vector field $\ell$. Therefore, the null flow determines an
evolution of tensors defined in the fibers of $L^\perp$.
Particularly important will be for us the tensor  $q_{ab}(x)$
induced in each fiber $L^\perp_x$ of $L^\perp$ by the restriction
of the space-time metric tensor $g_{\alpha\beta}(x)$. The induced
tensor  is often referred to as the degenerate  metric tensor.
Indeed, for every point $x\in\M$, $q_{ab}(x)$ is symmetric and
being defined in the $n-1$ dimensional fiber $L^\perp_x$, it
 has the signature $(0,n)$. The evolution  of the field $q_{ab}$ defined by
the null flow
is just
\begin{equation}\label{eq:lie_q}
  \lie_\ell q_{ab}\ =\ 2\Bvel_{(ab)} \ ,
\end{equation}
where  $\Bvel_{(ab)}(x)$ is the restriction of the derivative tensor
\begin{equation}\label{eq:B_def}
  \Bvel_{\alpha\beta}\ =\ \nabla_\beta\ell_\alpha \ ,
\end{equation}
to the vector space $L^\perp_x$, at each $x\in\M$. Note, that
\begin{subequations}\label{eq:Bl_contr}\begin{align}
  \ell^{\alpha}\Bvel_{\alpha\beta}\ &=\ 0 \ ,  &
  \ell^{\beta}\Bvel_{\alpha\beta}\ &=\ \sgr{\ell}\ell_{\alpha} \ ,
\tag{\ref{eq:Bl_contr}}\end{align}\end{subequations}
therefore the restriction $\Bvel_{ab}$ is annihilated by $\ell$,
\begin{subequations}\label{eq:Bl_contr_a}\begin{align}
  \ell^{a}\Bvel_{ab}\ &=\ 0 \ , &
  \ell^{b}\Bvel_{ab}\ &=\ 0 \ . \tag{\ref{eq:Bl_contr_a}}
\end{align}\end{subequations}
The null flow evolution of $\Bvel_{\alpha\beta}$ involves
the space-time Riemann tensor $R_{\alpha\beta\gamma\delta}$,
\begin{equation}\label{eq:DB}
  \lie_{\ell}\Bvel_{\alpha\beta}
  \ =\ \sgr{\ell} \Bvel_{\alpha\beta}
  + \ell_{\alpha}\sgr{\ell}{}_{,\beta}
  + \Bvel_{\gamma\alpha}{\Bvel^\gamma}_\beta
  - \Riem{n}_{\mu\alpha\nu\beta}{\ell}^\mu{\ell}^\nu \ .
\end{equation}
To read from this equation the equation of the null flow evolution
of $q_{ab}$ it is convenient to consider the quotient
bundle $L^\perp/L$, whose fiber at every $x\in\M$, is the quotient vector
space $L^\perp_x/L_x$ of the fibers. Given a covariant tensor
$\bar{C}_{A...B}$ in $L_x^\perp/L_x$, we denote its pullback to
$L_x^\perp$ by ${C}_{a...b}$; given a vector $X^a$ at $x$ orthogonal to
$L_x$ we denote by $\bar{X}^A$ its projection onto
$L_x^\perp/L_x$.  Examples are the very tensors $q_{ab}(x)$ and
$\Bvel_{ab}(x)$, who are in fact pullbacks of  tensors defined
in  $L_x^\perp/L_x$, consequently denoted by $\bar{q}_{AB}(x)$ and,
respectively,
$\bBvel_{AB}(x)$. The first of them, $\bq_{AB}$ is a non-degenerate,
positive definite metric tensor in each fiber of $L^\perp/L$.
The quotient bundle is also preserved by the null flow. Using
(\ref{eq:lie_q}) we can see, that
\begin{equation}
  \lie_\ell \bq_{AB}\ =\ 2\bBvel_{(AB)} \ .
\end{equation}
The tensor $\bBvel_{AB}$ can be decomposed into three parts:
\begin{itemize}
  \item The trace with respect to $\bq_{AB}$ (with $\bq^{AB}$ being an
        inverse of $\bq_{AB}$)
        \begin{equation} \label{eq:exp_l}
          \thv{\ell}\ :=\ \bq^{AB}\bBvel_{AB}  \ ,
        \end{equation} which is called the expansion scalar,
  \item the  traceless symmetric part:
        \begin{equation} \label{eq:shr_l}
          \sgv{\ell}{}_{AB}\
          :=\ \bBvel_{AB}
             - \fracs{1}{n-2}\thv{\ell}\bq_{AB} \ ,
        \end{equation}
        called the shear tensor
\item and the antisymmetric part $\bBvel_{[AB]}$ \ .
\end{itemize}

Since the transformation law for the tensor $\Bvel_{\alpha\beta}$ upon
rescalings
$\ell \mapsto \ell'=f\ell$ by a function $f$ is quite simple, namely
\begin{equation}\label{eq:B_trans'}
  \Bv{\ell'}_{\alpha\beta}\
  =\ f\Bvel_{\alpha\beta} + \ell_\alpha D_\beta f \ ,
\end{equation}
and in particular
\begin{subequations}\label{eq:B_trans}\begin{align}
  \Bv{\ell'}_{ab}\ &=\ f\Bvel_{ab} \ ,  &
  \bBv{\ell'}_{AB}\ &=\ f\bBvel_{AB} \ , \tag{\ref{eq:B_trans}}
\end{align}\end{subequations}
it is often convenient to choose a  section $\ell_o$ of the
bundle $L$, such that
\begin{equation}
  \sgr{\ell_o}\ =\ 0
\end{equation}
in \eqref{eq:geo}.
The evolution of the corresponding $\bBvlo_{AB}$ defined by
the flow of $\ell_o$ is
\begin{equation}\label{eq:DBAB}
  \lie_{\ell_o}\bBvlo_{AB}
  \ =\ \bq^{CD}\bBvlo_{CA}\bBvlo_{DB}
    - \bRiem{n}_{\alpha A\beta B}\ell_o^\alpha\ell_o^\beta \ .
\end{equation}
In particular,
\begin{equation}\label{eq:Ray}
  \lie_{\ell_o} \thv{\ell_o}\ 
  =\ - \fracs{1}{n-2}(\thv{\ell_o})^2
     - \sgv{\ell_o}_{AB}\sgv{\ell_o}^{AB} + \bBvlo_{[AB]}\bBvlo^{[AB]}
     - \ell_o{}^{\mu}\ell_o{}^{\nu}\Ricn{n}_{\mu\nu} \ ,
\end{equation}
where $\Ricn{n}_{\mu\nu}$ is the spacetime Ricci tensor and the
capital indexes are raised with the inverse metric tensor
$\bq^{AB}$. Note, that this geometric identity defines the
dynamics of the geometry $\bar{q}_{AB}$ if we use the Einstein
equations and replace the Ricci tensor by the matter
energy-momentum. This is a straightforward generalization of the
famous Raychaudhuri equation in 4-dimensional space-time. The
essential feature of this equation is still present in this $n$
dimensional case: all the terms on the right hand side except
$\bBvlo_{[AB]}\bBvlo^{[AB]}$ are non-positive, provided the
Einstein equations hold and the energy  condition
\begin{equation}\label{eq:Tll>0}
  T_{\alpha\beta}\ell^\alpha\ell^\beta\ \ge \ 0
\end{equation}
is satisfied by the matter. In particular, the non-negativity of
$\sgv{\ell}_{AB}\sgv{\ell}^{AB}$ follows from the positive definiteness
of the metric tensor field $\bq_{AB}$.

We have not exhausted all the information contained in the tensor
$\Bvel_{\alpha\beta}$ and in the equation \eqref{eq:DB}. We will
go back to them in the context of the Zeroth Law of the
non-expanding null surface thermodynamics.

\subsection{Null surfaces}
\label{sec:ns}

An $n-1$ dimensional submanifold $\ih$ in $\M$ is called a null
surface if at every point $x\in\ih$ the pullback $q_{ab}(x)$ of
the metric tensor $g_{\alpha\beta}(x)$ onto $\ih$ is degenerate.
Denote by $L_x$ the degeneracy subspace $L_x\subset T_x\ih$. It
follows from the algebra of a metric tensor of the signature
$(1,n-1)$ that $L_x$ is 1-dimensional at each point $x$, provided
the $g_{\alpha\beta}(x)$ is non-degenerate. It
consists of the null vectors tangent to the surface $\ih$ at $x$.
The spaces $L_x$ form a sub-bundle $L\subset T\ih$ referred to
through out this paper as {\it the null direction bundle}.
Consider an arbitrary null vector field $\ell^a$ defined (locally)
on $\ih$, a (local) section of the bundle $L$. It is geodesic,
that is it satisfies (\ref{eq:geo}). The function $\sgr{\ell}$ is
referred to as the surface gravity corresponding to $\ell$. To
apply the definitions and results of Section \ref{sec:ngfR} (in
particular the Raychaudhuri equation) to the vector field $\ell^a$
defined on $\ih$ it is enough to consider at each point $x\in\ih$
an appropriate local extension of the bundle $L$ and of the vector
field $\ell^a$ to a neighborhood of $x$ in $\M$. Such extension
always exists. Obviously, the tensor
$\Bvel_{\beta\alpha}=\nabla_\alpha \ell_\beta$ depends on the
extension, however, at the surface $\ih$, the part $\Bvel_{\beta
a}$ defined by the restriction of the derivative to the tangent
space to $\ih$ is the extension independent. Moreover, due to
(\ref{eq:Bl_contr}) the tensor $\Bvel^\beta{}_a$ considered as a
vector valued 1-form  defined on $\ih$ takes values in the tangent
bundle $T\ih$, therefore it is defined intrinsically on $\ih$ and
can be denoted by $\Bvel^b{}_a$. Another object defined
intrinsically on $\ih$ is $\lie_\ell \Bvel^\beta{}_a$. In this way
the equations (\ref{eq:DB}, \ref{eq:DBAB}, \ref{eq:Ray})
 can be applied to every null vector field
$\ell^a$ defined on and tangent to $\ih$. The equations describe the
evolution of the tensors $q_{ab}$ and $B^b{}_a$ along $\ih$,
defined by $\ell^a$. The existence of the surface implies that the
antisymmetric part of the pullback $\Bvel_{ab}$ vanishes,
\begin{equation}\label{eq:Bant}
  \Bvel_{[ab]}\ =\ 0 \ .
\end{equation}
To see this, owing to \eqref{eq:B_trans}, it suffices to show \eqref{eq:Bant}
for arbitrary one non-trivial example of $\ell^a$. Consider a function $r$
defined in a neighborhood of a point of $\ih$ in $\M$ such that
\begin{subequations}\label{eq:def_r}\begin{align}
  r|_{\ih}\ &=\ \const \ , &
  \rd r|_{\ih}\ &\neq\ 0 \ . \tag{\ref{eq:def_r}}
\end{align}\end{subequations}
Then $\ell_{o\mu}=\nabla_\mu r$ defines a vector field
$\ell_{o}^a$ tangent to the surface $\ih$ and null thereon. $\Bvlo_{ab}$ is
the pullback to $\ih$ of the symmetric space-time tensor
$\nabla_\beta\nabla_\alpha r$, so it is symmetric itself.
In the consequence of \eqref{eq:Bant}, the Raychaudhuri equation reads
\begin{equation}\label{eq:Rayns}
  \lie_{\ell_o} \thv{\ell_o}\
  =\ - \fracs{1}{n-2}(\thv{\ell_o})^2
    - \sgv{\ell_o}_{AB}\sgv{\ell_o}^{AB}
    - \ell_o{}^{\mu}\ell_o{}^{\nu}\Ricn{n}_{\mu\nu} \ ,
\end{equation}
in the case of $\ell^a=\ell_o^a$ such that the corresponding surface gravity
$\sgr{\ell_o}$ vanishes.

\section{Non-expanding null surfaces}
\label{sec:nes}

\subsection{Definition}
\label{sec:nes_def}

Suppose, that given a null surface $\ih$, for every point $x\in\ih$ the
expansion $\thv{\ell}$ of some non-trivial null vector field $\ell^a$ tangent
to $\ih$ at $x$ vanishes,
\begin{equation}
  \thv{\ell}\ =\ 0 \ .
\end{equation}
 Then, we say that $\ih$ is non-expanding.
This is a property of the surface $\ih$  only, independent of a
choice of $\ell$. Indeed, it follows from \eqref{eq:B_trans}, that
if at a given point $x$  the expansion $\thv{\ell}$ vanishes, then
the same is true for every other section $\ell'$ of the bundle
$L$.

\subsection{The vanishing of the shear and  $\Ricn{n}_{\ell\ell}$}

To learn more about the non-expanding null surface case, consider again a
vector field $\ell_o^a$, a section of the bundle $L$, such that
\begin{equation}
  \sgr{\ell_o}\ =\ 0 \ .
\end{equation}
The vanishing of the left
hand side (the expansion $\thv{\ell_o}$) of \eqref{eq:Rayns}
and  the vanishing of the antisymmetric part of $\bBvlo_{AB}$ lead to
\begin{equation}
  0\ \ =\ \sgv{\ell_o}_{AB}\sgv{\ell_o}^{AB}\
       +\ \ell_o{}^{\mu}\ell_o{}^{\nu}\Ricn{n}_{\mu\nu} \ .
\end{equation}
The first term on the right hand side is non-negative. We assume the energy
inequality \eqref{eq:Tll>0} which makes the second term also
non-negative. Hence,
all of them necessarily vanish on $\ih$. Moreover, it follows from the
positivity of the metric tensor $\bq_{AB}$ and the symmetry of
$\sgv{\ell}_{AB}$, that
\begin{equation}
  \sgv{\ell_o}_{AB}\sgv{\ell_o}^{AB}\ =\ 0 \
  \Rightarrow\
  \sgv{\ell_o}_{AB}\ =\ 0 \ .
\end{equation}
In the consequence,
\begin{equation}
  \Bvlo_{ab} \ =\ 0\
  =\ \ell_o{}^{\mu}\ell_o{}^{\nu}\Ricn{n}_{\mu\nu} \ ,
\end{equation}
on $\ih$.

Since the tensor $\bBvel_{ab}$ transforms as presented in \eqref{eq:B_trans}
the final conclusion is true for arbitrary choice of a section $\ell$ of
the bundle $L$.

\begin{theorem}\label{thm:nes}
  Suppose $\ih$ is a non-expanding, null, $n-1$ dimensional surface contained
  in a  space-time of signature $(1,n-1)$; suppose
  the Einstein field equation hold on $\ih$ with a
  cosmological constant and with
  the matter fields which satisfy the energy condition
  \eqref{eq:Tll>0}. Then:
  \begin{enumerate}[ (i)]
    \item the surface is shear-free, that is for
      every null vector field $\ell^\alpha$ defined on and tangent to $\ih$
      \begin{equation}
        \nabla_a\ell_b\ =\ 0 \ ,
      \end{equation}
      where $\nabla_a\ell_b$ is the pullback of the spacetime
      $\nabla_\alpha\ell_\beta$ to $\ih$;
    \label{it:q_flow}
    \item the induced degenerate metric $q_{ab}$ in $\ih$
      is invariant with respect to the flow of every null vector field
      $\ell^a$ tangent to $\ih$,
      \begin{equation}\label{eq:nes_q}
        \lie_\ell q_{ab}\ =\ 0 \ ,
      \end{equation}
    \item The space-time Ricci tensor satisfies on $\ih$ the
      following  condition
      \begin{equation}\label{eq:shFree}
        \Ricn{n}_{\alpha\beta}\ell^{\alpha}\ell^{\beta}\ =\ 0 \ .
      \end{equation}
  \end{enumerate}
\end{theorem}

The property \eqref{it:q_flow} above combined with  $\ell^aq_{ab}=0$ means that,
locally $q_{ab}$ is the pullback of certain metric tensor field $\hq_{AB}$
defined on an $n-2$-dimensional manifold $\bas'$. The manifold $\bas'$ is the
set of the null curves tangent to the null direction bundle $L$ in appropriate
neighborhood $\ih'\subset\ih$ open in $\ih$, and the map is the natural
projection,
\begin{subequations}\label{eq:bas}\begin{align}
  \pi:\ih'\ &\rightarrow\ \bas' \ , &
  q_{ab}\ &=\ \pi^*\hq_{AB} \ . \tag{\ref{eq:bas}}
\end{align}\end{subequations}

\subsection{The induced covariant derivative}

If the assumptions of Theorem \ref{thm:nes}  are satisfied, then
for any vector fields $X,Y$, sections of the tangent bundle
$T\ih$, the covariant derivative $\nabla_XY$ is again a vector
field tangent to $\ih$. Indeed, it is easy to see that
$\nabla_XY$ is necessarily orthogonal to $\ell$,
\begin{equation}
  \ell_{\mu}X^{\nu}\nabla_{\nu}Y^{\mu}\ 
  =\ X^{\nu}\nabla_{\nu}\left(\ell_{\mu}Y^{\mu}\right)
  - X^{\nu}Y^{\mu}B_{\nu\mu} = 0 \ ,
\end{equation}
where the first term vanishes by the definition of $\ell$ and the second
due to Theorem \ref{thm:nes}.
The induced covariant derivative  will be denoted by $D_a$.
For the vector fields $X^a,Y^a$, sections of the bundle $T\ih$,
the derivative is just
\begin{equation}\label{eq:D_def}
  D_XY^a := \nabla_XY^a \ ,
\end{equation}
whereas for a covector $W_a$, a section of the dual bundle $T^*\ih$,
the derivative $D_XW_a$ is determined by the Leibnitz rule,
\begin{equation}
  Y^aD_XW_a\ =\ D_X(Y^aW_a) - (D_XY^a)W_a \ .
\end{equation}
Obviously, the derivative $D_a$ is torsion free and annihilates
the degenerate metric tensor $q_{ab}$,
\begin{subequations}\label{eq:D-tfree}\begin{align}
  D_aD_b f\ &=\ D_bD_a f \ ,  &
  D_aq_{bc}\ &=\ 0 \ ,  \tag{\ref{eq:D-tfree}}
\end{align}\end{subequations}
for every function $f$.

\subsection{Further  conditions on the Riemann tensor necessary
at $\ih$}
\label{sec:fur_Riem}

The conclusions of Theorem \eqref{thm:nes} lead to stronger
restrictions on the Riemann tensor at $\ih$, namely
\begin{equation}\label{eq:Riem_labc}
  \Riem{n}{}_{abc\alpha}\ell^\alpha\ =\ 0
\end{equation}
where $\ell^a$ is a null vector tangent to $\ih$.
Indeed,  the contraction of the Riemann tensor with $\ell$
and any vector fields $X^a,Y^a,Z^a$, sections of $T\ih$, can be expressed
as a functional homogeneous in the derivative tensor $\Bvel_{ab}$:
\begin{equation}\label{eq:R_tang}\begin{split}
  X^{\mu}\ell^{\nu}Y^{\alpha}Z^{\beta}\Riem{n}_{\mu\nu\alpha\beta}\
  &=\ X^{\mu}Y^{\alpha}Z^{\beta}\nabla_{[\alpha}\nabla_{\beta]}\ell_{\mu} \\
  &=\ Y^{a}D_{a}X^{m}Z^{b}\Bvel_{bm}
     - \Bvel_{b m} ( X^{m}Y^{a}D_{a}Z^{b}
                         + Z^{b}Y^{a}D_{a}X^{m} ) \\
  &\ - Z^{b}D_{b}X^{m}Y^{a}\Bvel_{am}
     + \Bvel_{am} ( X^{m}Z^{b}D_{b}Y^{a}
                      + Y^{a}Z^{b}D_{b}X^{m} ) \ .
\end{split}\end{equation}
Note, that in the calculation we have used the fact, that the
space-time covariant derivative applied in any direction tangent
to $\ih$ preserves the tangent bundle $T\ih$.

Thus far only the inequality
$T_{\alpha\beta}\ell^\alpha\ell^\beta\ge 0$ was used apart from
the zero expansion assumption and the Einstein equations with
possibly non-zero cosmological constant. A somewhat stronger but
still quite mild assumption about the energy-momentum tensor
$T_{\alpha\beta}$ is\footnote{
  This inequality automatically holds when the dominant energy condition
  is assumed, but is much weaker. The consequences of dropping of the
  condition  will be briefly discussed in the
  appendix.}:

\begin{cond}\label{c:energy}(Stronger Energy Condition)
  At every point of the surface $\ih$, the vector field
  \begin{equation}
    -T^{\mu}{}_{\nu}\ell^{\nu}
  \end{equation}
  is causal, that is
  \begin{equation}\label{eq:T_cau}
    g^{\mu\nu}T_{\mu\alpha}\ell^{\alpha}T_{\nu\beta}\ell^{\beta}\ \leq\ 0 \ ,
  \end{equation}
  and future oriented, for every future oriented null vector
  field $\ell$ defined on and tangent to $\ih$.
\end{cond}
\noindent This condition implies in particular, the previous
\begin{equation}
  T_{\ell\ell}\ \geq\ 0 \ .
\end{equation}
\medskip
Now, the vanishing of the Ricci tensor component $
\ell^{\mu}\ell^{\nu}\Ricn{n}_{\mu\nu}$ on $\ih$ combined with
Stronger Energy Condition \eqref{c:energy} leads to further
restrictions on the Ricci tensor. Consider the 1-form
\begin{equation}
  \Ricl_a\ :=\ \Ricn{n}_{a\beta}\ell^{\beta}\ ,
\end{equation}
a section of the cotangent bundle $T^*\ih$. Due to the vanishing
of $\Ricn{n}_{\ell\ell}$, at each $x\in \ih$, $\Ricl_a\in
T^*_x\ih$ is the pullback of some $\bRicl_A\in (T_x\ih/L)^*$. The
Einstein field equations allow us to express the non-positive
space-time norm of the  field $T^{\mu}{}_{\nu}\ell^{\nu}$ by the
non-negative norm of $\bRicl_A$ with respect to  $\bar{q}^{AB}$,
\begin{equation}\label{eq:TRnorm}
  0\ \geq\  g^{\mu\nu}T_{\mu\alpha}\ell^{\alpha}T_{\nu\beta}\ell^{\beta}\
     =\ \bq^{AB} \bRicl_A \bRicl_{B}\ \geq \ 0\ .
\end{equation}
Hence, the pullback onto $\ih$ of the Ricci tensor contracted with $\ell$
is identically zero at $\ih$,
\begin{equation}\label{eq:Ric_la}
  \Ricn{n}_{a\beta}\ell^{\beta}\ =\ 0\ .
\end{equation}

Combining this result with the condition \eqref{eq:Riem_labc} on
the Riemann tensor one can obtain the following condition on  the
space-time Weyl tensor at the NEH:
\begin{equation}\label{eq:Weyl}
  \Weyl{n}{}_{abc\delta}\ell^{\delta} |_{\ih}\ =\ 0 \ .
\end{equation}
In $n=4$ case the condition  means that the null direction tangent
to the surface $\ih$  is a double principal null direction of the
Weyl tensor. In \cite{Pravda} the Petrov classification of the
Weyl tensor was generalized to an arbitrary dimension. The Weyl
tensor was expressed in a frame built of  real vectors
$(n,\ell,\theta_{(A)})$ such that\footnote{The proposed frame is
  an analog of the Newman-Penrose complex null tetrad
  \cite{exact}. The pair of complex null vectors $(m,\bar{m})$ is
  replaced by a set of real spacelike unit vectors which allows to
  generalize the frame to arbitrary dimension.}
\begin{itemize}
  \item On a given n dimensional manifold $\M$ the the vectors
         $(n,\ell)$ are null and normalized by the condition
         $\ell^{\mu}n_{\mu}=-1$.
  \item The spacelike vectors $\theta_{(A)}$ constitute the orthonormal
         basis of the subspace of $T\M$ orthogonal to $(n,\ell)$.
\end{itemize}
The proposed classification is based on the behavior of the Weyl
tensor under the boost transformations $(\ell\mapsto f\ell,
n\mapsto f^{-1}n)$. For a given (fixed) $\ell$ the Weyl tensor can
be decomposed onto the sum of the terms $C^{(b)}$, such that each
of them transforms under the boost in the following way:
$C^{(b)}\mapsto f^bC^{(b)}$. The integer power $b$ is called the
boost weight. The weight of the leading term (denoted as the boost
order $\mathcal{B}(\ell)$) depends on the $\ell$ only. Therefore
for a given Weyl tensor one can distinguish the set of {\it
aligned} vectors of the boost order $\mathcal{B}(\ell)\leq 1$. The
Weyl tensor is classified as being of the type I (II,III,N) if
there exists null vector $\ell$ of the boost order $1$
($0,-1,-2$), and there does not exist a null vector of a lower
order. The condition \eqref{eq:Weyl} implies that the boost order
of the null direction tangent to $\ih$ is at most $0$, so the Weyl
tensor is at least of the type II with respect to the principal
classification introduced in \cite{Pravda} and sketched above.

\subsection{Rotation}
\label{sec:oth}

The covariant derivative $D_a$  induced on $\ih$  preserves the null direction
bundle $L$. Indeed, for every  section  $\ell^a$  of $L$,
and every vector field $X^a$, a section of $T\ih$, the vector field $D_X\ell^a$
is orthogonal to every vector $Y^a$ tangent to $\ih$,
\begin{equation}
  q_{ab}Y^{a}D_X\ell^{b}\ 
  =\ - q_{ab}\ell^{b}D_{c}Y^{a}\ 
  =\ 0 \ . 
\end{equation}
That implies that the derivative $D_a\ell^b$ is proportional to $\ell^b$
itself,
\begin{equation}\label{eq:omega_def}
  D_a\ell^b\ =\ \Bvel{}_a{}^b\ =\  \w{\ell}{}_a\ell^b \ ,
\end{equation}
where  $\w{\ell}{}_a$ is a 1-form defined uniquely on this subset
of $\ih$ on which $\ell\neq0$ is defined. We call $\w{\ell}{}_a$
the rotation 1-form potential, as a generalization of the $n=4$
dimensional  case \cite{abl-g}.

In $4$ dimensions, the evolution of $\w{\ell}{}_a$ along the
surface $\ih$ upon the null flow is responsible for the Zeroth Low
of the non-expanding horizon thermodynamics. Therefore, we study
this equation in the current case. It is convenient to investigate
the behavior of the following object:
\begin{equation}
  \ell^b\lie_{\ell}\w{\ell}{}_a\ =\ \lie_{\ell} \Bvel{}^b{}_{a} \ .
\end{equation}
The right hand side is given by \eqref{eq:DB}, and after a short calculation
it reads as follows,
\begin{equation}\label{eq:lieomega}
  \ell^b\lie_{\ell}\w{\ell}{}_a\
  =\ \ell^b D_a\sgr{\ell} - \Riem{n}_c{}^b{}_{da}\ell^c\ell^d \ .
\end{equation}
where we used the fact that
\begin{equation}
  \sgr{\ell}\ =\ \w{\ell}{}_a\ell^a \ .
\end{equation}

The vanishing of the components $\Riem{n}_{abcd}\ell^d$
(see \eqref{eq:R_tang})
allows us to express the Riemann tensor component appearing in
\eqref{eq:lieomega} by the Ricci tensor
\begin{equation}
  \Riem{n}_c{}^b{}_{da}\ell^c\ell^d\ =\ - \Ricn{n}_{ca}\ell^b\ell^c \ ,
\end{equation}
hence the vector field $\ell^a$ can be completely factored out,
\begin{equation}\label{eq:pre0th}
  \lie_{\ell}\w{\ell}{}_a\ =\ D_a\sgr{\ell} \ +\ \Ricn{n}_{ab}\ell^b
\end{equation}

If Stronger  Energy Condition \eqref{c:energy} holds then  the
last term above also vanishes (see Section \ref{sec:fur_Riem}).

In conclusion, the evolution of the rotation potential is described by the
following theorem:
\begin{theorem}[The Zeroth Law]\label{thm:0th}
  Suppose $\ih$ is an $n-1$ dimensional, non-expanding, null
  surface; suppose
  that the assumptions of Theorem \ref{thm:nes} and Stronger
  Energy Condition
  \ref{c:energy} are satisfied.
  Then, for every null vector field $\ell^a$ defined on and tangent to
  $\ih$, the corresponding rotation 1-form potential $\w{\ell}{}$ and the
  surface gravity $\sgr{\ell}$ satisfy the following constraint:
  \begin{equation}\label{eq:0th}
    \lie_{\ell}\w{\ell}{}_a\ =\ D_a\sgr{\ell} \ .
  \end{equation}
\end{theorem}

Theorem \ref{thm:0th} tells us, that there is always a choice of
the section $\ell$ of the null direction bundle $L$ such that
$\w{\ell}{}$ is Lie dragged by $\ell$. For,  we can always find a
non-trivial section $\ell$ of $L$ such that $\sgr{\ell}$ is
constant. The relation with the original Zeroth Law of black hole
thermodynamic goes the other way around. Indeed, if the vector
field $\ell^a$ admits an extension to a Killing vector defined in
a neighborhood of $\ih$, then $\w{\ell}{}$ is Lie dragged by the
flow, therefore the left hand side is zero, hence $\sgr{\ell}$ is
necessarily (locally) constant.

The dependence of the rotation 1-form potential $\w{\ell}{}_a$
on a choice  of the section $\ell^a$ of $L$ follows from \eqref{eq:B_trans'}:
If ${\ell'}^a=f\ell^a$, then
\begin{equation}\label{eq:omega'}
  \w{\ell'}{}_a\ =\ \w{\ell}{}_a + D_a \ln f \ .
\end{equation}
As one can see, its external derivative (in the sense of the
manifold $\ih$) is the surface $\ih$ invariant,
\begin{equation}\label{eq:Omega}
  \Omega_{ab}\ :=\ D_a\w{\ell}{}_b - D_b\w{\ell}{}_a\
    =\ D_a\w{\ell'}{}_b - D_b\w{\ell'}{}_a \ .
\end{equation}
We call it the rotation 2-form. Note, that whereas the sections of
the null direction bundle $L$ were considered on $\ih$ locally,
and so were the corresponding rotation 1-form potentials
$\w{\ell'}{}_a$, the rotation 2-form is defined globally on $\ih$.
An immediate consequence of Theorem \eqref{thm:0th} is that
whenever the assumptions are satisfied, the rotation 2-form is
orthogonal to the bundle $L$, and Lie dragged by any (local) null
flow defined by a section $\ell$ of $L$,
\begin{subequations}\begin{align}
  \ell^a\Omega_{ab}\ =\ \lie_\ell\w{\ell}_b - D_b(\ell^a\w{\ell}_a)\
    &=\ 0\ , \\
  \lie_\ell \Omega_{ab}\ &=\ 0 \ .
\end{align}\end{subequations}
Therefore, the rotation 2-form $\Omega_{ab}$ is at every point $x\in\ih$ the
pullback with respect to $T_x\ih\rightarrow T_x\ih/L_x$ of a tensor
$\bar{\Omega}_{AB}$ defined in $T_x\ih/L_x$, and such that
\begin{equation}\label{eq:LOmega}
  \lie_\ell\bar{\Omega}_{AB}\ =\ 0 \ .
\end{equation}

\subsection{Geometry and the constraints}
\label{sec:D_evo} 

Given a non-expanding null surface $\ih$, the
pair $(q_{ab}, D_a)$, that is the induced degenerate metric and,
respectively, the induced covariant derivative are referred to as
the geometry of $\ih$.  By a `constraint' on the non-expanding
surface geometry we mean here every geometric identity $F(q_{ab},
D_a,\Ricn{n}_{\alpha\beta})=0$ involving the geometry
$(q_{ab},D_a)$ and the space-time Ricci tensor at $\ih$ only. Part
of the constraints is already solved by  Theorem \ref{thm:nes}
$(ii)$, that is by the conclusion that $q_{ab}$ be Lie dragged by
every null flow generated a null vector field $\ell$ tangent to
$\ih$. Another example of a constraint is the Zeroth Law
(\ref{eq:pre0th}, \ref{eq:0th}). A complete set of the
functionally independent constraints is formed by $\lie_\ell
q_{ab}=0$ and by an identity satisfied by the commutator
$[\lie_\ell,D_a]$, where $\ell$ is a fixed, non-vanishing section
of the null direction bundle $L$. We turn now to the second
identity mentioned above.

Using the formula \eqref{eq:omega_def}, and using
\eqref{eq:Riem_labc}, after simple calculations  one can express
the value of the commutator $[\lie_\ell,D_a]$ by the rotation
potential, its derivative, and the space-time Riemann tensor,
\begin{equation}
  [\lie_\ell,D_a]X^b\ =\ \left[ \ell^b\left(D_{(a}\w{\ell}{}_{c)}\
    +\ \w{\ell}{}_a\w{\ell}{}_c\right)
    -\ \ell^\delta \Riem{n}{}^b{}_{(ca)\delta} \right]X^c
\end{equation}
(where $D_a\w{\ell}{}_c$, stands for the tensor, not for a
derivative operator acting on $X^b$). It follows from the
condition $\eqref{eq:Riem_labc}$, that
$\ell^\delta\Riem{n+2}{}^b{}_{(ca)\delta}$ is also proportional to
$\ell^b$, and we can write
\begin{equation}\label{eq:[l,D]1}
  [\lie_\ell,D_a]X^b\ =\ \ell^b N_{ac}X^c \ .
\end{equation}
To spell out what the proportionality factor $N_{ac}$ is we need
to remind that the degenerate metric tensor field $q_{ab}$ can be
locally defined as the pullback \eqref{eq:bas} of the
$n$-dimensional metric tensor $\hq_{AB}$ defined on the manifold
of the curves generating $\ih$. The proportionality factor can be
expressed by the pullback spacetime Ricci tensor and the
pullback\footnote{The first pullback is defined by the embedding
$\ih\rightarrow \M$, whereas the second one corresponds to the
locally defined projection $\pi:\ih'\rightarrow \ \hat{\ih}'$ of a
neighborhood $\ih'\subset \ih$ onto the space of the null curves
in $\ih'$.} of the Ricci tensor $\Ricn{n-2}_{AB}$ of the metric
$\hq_{AB}$, namely
\begin{equation}\label{eq:[l,D]2}
  N_{ac}\ =\  D_{(a}\w{\ell}{}_{c)}\ +\ \w{\ell}{}_a\w{\ell}{}_c
  + \frac{1}{2}\left(\Ricn{n}_{ac}-\pi^*\Ricn{n-2}_{ac}\right) \ .
\end{equation}
The identities (\ref{eq:[l,D]1}, \ref{eq:[l,D]2}) are the
constraints in the sense explained at the beginning of this
subsection. They become the gravitational part of the genuine
Einstein constraints when the space-time Ricci tensor is replaced
by the cosmological constant part and by the energy momentum
tensor of the matter field. As an example, later we will consider
the vacuum case. We did not assume  Stronger Energy Condition
\eqref{c:energy} to derive (\ref{eq:[l,D]1}, \ref{eq:[l,D]2}).

The contraction of (\ref{eq:[l,D]1}, \ref{eq:[l,D]2}) with
$\ell^a$ is equivalent to (\ref{eq:pre0th}). Hence it defines the
evolution of the rotation 1-form potential $\w{\ell}_a$ already
used in the proof of the Zeroth Law. Recall, that locally there is
on $\ih$ a nowhere vanishing tangent, null vector field $\ell_o$
such that
\begin{equation}
  \sgr{\ell_o}\ =\ 0 \ .
\end{equation}
The corresponding $\w{\ell_o}$ is Lie dragged by the vector field
$\ell_o$, provided the assumptions of the Zeroth Law are
satisfied,
\begin{equation}
  \lie_\ell \w{\ell_o}\ =\ 0 \ .
\end{equation}
The meaning of the remaining part of the
constraint (\ref{eq:[l,D]1}, \ref{eq:[l,D]2}) is explained in the
next sub-section after we itemize the derivative $D_a$ into
components and provide a more explicit form of eqs.
(\ref{eq:[l,D]1}, \ref{eq:[l,D]2}).

\section{Elements of the non-expanding null surface geometry}
\label{sec:elements}

\subsection{Compatible coordinates, foliations}
\label{sec:slices}

To understand better the elements of the covariant derivative
$D_a$ induced on a null, non-expanding surface $\ih$, and to
investigate further its relation with the space-time Ricci tensor,
we need to introduce an extra local structure on $\ih$.

Let  $\ell^a$ be a nowhere vanishing local section of the null
direction bundle $L$. Given $\ell^a$,  let $v$ be a real function
defined in the domain of $\ell^a$, compatible with $\ell^a$, that
is such that
\begin{equation}\label{eq:def_v}
  \ell^a D_a v\ =\ 1 \ .
\end{equation}
the function $v$ exists provided we sufficiently reduce the domain
of $\ell^a$. We will refer to $v$ as to a coordinate compatible
with $\ell$. The function $v$ is used to define a covector field
on $\ih$,
\begin{equation}
  n_a\ :=\ -D_av \ .
\end{equation}
The covector field has the following properties,
\begin{enumerate}[ (i)]
  \item \label{it:n_norm}
        Is normalized in the sense that
        \begin{equation}\label{eq:n_norm}
          \ell^a n_a\ =\ -1 \ ,
        \end{equation}
  \item \label{it:n_fol}
        It is orthogonal to the constancy surfaces $\slc_v$ of the
        function $v$. 
\end{enumerate}
The surfaces $\slc_v$ will be referred to as slices. The family of
he slices  is preserved by the null flow of $\ell$, and so is $n_a$,
\begin{equation}
  \lie_\ell n_a\ =\ 0 \ .
\end{equation}

At every point $x\in \ih$,  the tensor
\begin{equation}\label{eq:g_dec}
  \tq^{a}{}_{b}\ :=\ \delta^{a}{}_{b} + \ell^{a}n_{b}
\end{equation}
defines the orthogonal to $\ell^a$ projection
\begin{equation}
  T_x\ih\ \ni\ X^a\ 
  \mapsto\  \tilde{X}^a\ =\ \tq^{a}{}_{b}X^b\ \in\ T_x\slc_v \ .
\end{equation}
onto the tangent space $T_x\slc_v$, where $\slc_v$ is the
slice passing through $x$. Hence, instead of $\tilde{X}^a$
we will write $\tilde{X}^A$, according to the index notation
explained in Introduction. Applied to the covectors, elements of
$T^*_xM$, on the other hand, $\tq^{a}{}_{b}$ maps each of them
into the pullback onto $\slc_v$,
\begin{equation}\label{tilde}
  T_x^*\ih\ \ni\ Y_a\ \mapsto\
  \tilde{Y}_a\ :=\ \tq^{b}{}_{a}Y_b\ \in\ T^*_x\slc_v \ ,
\end{equation}
hence the result will be also denoted by by using a capital Latin
index, as for example $\tilde{Y}_A$.

The field $n_a$ could be  extended to a section of the pullback
$T_\ih^*\M$ to $\ih$ of the cotangent bundle $T^*M$,  by the
requirement that
\begin{equation}\label{eq:n_null}
  g^{\mu\nu}n_{\mu}n_{\nu}\ =\ 0 \ .
\end{equation}
Hence $n_a$ can be thought of as a transversal to $\ih$ null
vector field from the space-time point of view.

\subsection{The elements of $D_a$}
\label{sec:D_comp}

Each slice $\slc_v$ of the foliation introduced above is equipped with the
induced metric
tensor $\tq_{AB}$ defined by the pullback of $q_{ab}$ (and of
$g_{\alpha\beta}$) to $\slc_v$.
Denote by $\tD_A$ the torsion free and metric  covariant derivative
determined on $\slc_v$ by the metric tensor $\tq_{AB}$.
All the slices are naturally isometric.

The covector field $n_a$ gives rise to the following symmetric tensor
defined on $\ih$,
\begin{equation}\label{eq:S_def}
  S_{ab}\ :=\ D_an_b \ .
\end{equation}

Given the  structure introduced  in the previous subsection
on $\ih$ locally (the null vector
field $\ell^a$, the foliation by slices $\slc_v$ and the covector field
$n_a$), the  derivative $D_a$ defined on $\ih$  is determined by the following
information
\begin{itemize}
  \item the torsion free covariant derivative $\tD_A$ corresponding
  to the Levi-Civita connection of the induced metric tensor
  $\tq_{AB}$ ,
  \item the rotation 1-form potential $\w{\ell}_a$, and
  \item a symmetric tensor $\tS_{AB}$ defined in each slice $\slc_v$,
    by the pullback of $D_an_b$,
    \begin{equation}\label{eq:SAB}
      \tS_{AB} \ =\ \tq^a{}_A\tq^b{}_BS_{ab} \ , \\
    \end{equation}
    and referred to the transversal expansion-shear tensor.
\end{itemize}
Indeed, for  every vector field $X^a$ and every covector field $Y_a$,
the sections of $T\ih$ and, respectively, $T^*\ih$, their derivative
can be composed from the following pieces:
\begin{subequations}\label{eq:D_dec}\begin{align}
   \tq^{a}{}_{A}\tq^{B}{}_{b}D_{a}X^{b}\
     &=\ \tD_{A}\tilde{X}^{B} \\
   \tq^{a}{}_{A}n_{b}D_{a}X^{b}\ 
     &=\ \tD_A(X^{b}n_{b}) - \tq^{a}{}_{A}S_{ab} \\
   \ell^{a}D_{a}X^{b}\ 
     &=\ \lie_{\ell}X^{b} + X^{a}\w{\ell}{}_{a}\ell^{b}  \\[0.2cm]
  \tq^a{}_A\tq^b{}_B D_a Y_b\ &=\ \tD_A \tilde{Y}_B
     - (Y_b\ell^b)\tq^a{}_A\tq^b{}_B S_{ab}\\
  \tq^a{}_A\ell^b D_a Y_b\ 
     &=\ \tD_A (\ell^b Y_b) - \tw{\ell}_A \ell^b Y_b \\
  \ell^a D_a Y_b\ &=\ \lie_{\ell}Y_b - \w{\ell}{}_b Y_a \ell^a
\end{align}\end{subequations}
where we have used the notation introduced in the previous section:
$\tilde{X}^A = \tq^{A}{}_{a}X^a,\ \ \tilde{Y}_A = \tq^{a}{}_{A}Y_a$,
$\tw{\ell}{}_A = \tq^{a}{}_{A}\w{\ell}{}_a$.

The careful reader noticed that all the components of the tensor
$S_{ab}$ were used above, not only the $\tS_{AB}$ part. However,
due to the normalization \eqref{eq:n_norm} the contraction of the
tensor with the null normal to $\ih$ is equal to:
\begin{equation}\label{eq:Sl}
  \ell^a S_{ab}\ =\ \w{\ell}{}_b \ .
\end{equation}

\subsection{The constraints on the elements of $D_a$}
\label{sec:wS_evo}

The constraints satisfied by $D_a$ are  expressed in the previous section
by the commutator $[\lie_\ell, D_a]$ (\ref{eq:[l,D]1},\, \ref{eq:[l,D]2}). Since
the foliation  we used to decompose $D_a$ into the elements $\tD_A$,
$\w{\ell}_a$ and $\tS_{AB}$
is invariant with respect to the flow of $\ell$, the evolution
of $D_a$ comes down to an evolution of $\tD_A$, $\w{\ell}_a$ and $\tS_{AB}$.
The slice connection $\tD_A$ is just invariant with respect to the flow,
because of  \eqref{eq:nes_q}.  The evolution of $\w{\ell}_a$ is already given
by the Zeroth Law \eqref{eq:0th}.  To describe the evolution of the remaining
element $\tS_{AB}$ we calculate the action of the commutator on the covector
$n_a$ and find, that the tensor $N_{ab}$ defined in ({\ref{eq:[l,D]1})
can be expressed by $\lie_{\ell}S_{ab}$, namely
\begin{equation}\label{eq:com_form}
  N_{ab}\ =\ \lie_{\ell}S_{ab} \ .
\end{equation}
Therefore, by \eqref{eq:[l,D]2},
\begin{equation}\label{S_evo_full}
  \lie_{\ell}S_{ab}\ =\ D_{(a}\w{\ell}_{b)} + \w{\ell}_a\w{\ell}_b
                    - \fracs{1}{2}\Riem{n}_{c(ab)}{}^d\ell^c n_d \ .
\end{equation}
The contraction of the expression with $\ell^a$ reproduces the Zeroth Law.
The more interesting at this point component, namely
the pullback of the tensor $\lie_{\ell}S_{ab}$ onto a slice $\slc_v$,
gives us the formula for the evolution of the transversal expansion-shear
tensor $\tS_{AB}$, we are asking about,
\begin{equation}\label{eq:S_ev}
  \lie_{\ell}\tS_{AB}\ =\ -\sgr{\ell}\tS_{AB} + \tD_{(A}\tw{\ell}_{B)}
    + \tw{\ell}_A\tw{\ell}_B - \fracs{1}{2}\,\Ricn{n-2}{}_{AB}
    + \fracs{1}{2}\tRicn{n}{}_{AB} \ ,
\end{equation}
where tilde consequently means the projection (\ref{tilde}), and
$\Ricn{n-2}{}_{AB}$ is the Ricci tensor of the metric tensor
induced in  slice $\slc_v$  (since locally, every slice $\ih_v$ is
naturally isometric with the space of the null curves $\hat{\ih}'$
equipped with the metric tensor $\hat{q}_{AB}$ we denote the
corresponding Ricci tensors in the same way).

\section{Isolated null surfaces}
\label{sec:iso}

\subsection{Definition, assumptions, constraints}
\label{sec:IHdef}

In this section we are continuing the study of the non-expanding,
null surfaces. We are  assuming that the Einstein equations with a
(possibly zero) cosmological constant hold on the surface, with
the energy-momentum tensor $T_{\alpha\beta}$ which satisfies
Stronger Energy Conditions \eqref{c:energy}. As it was shown,
these assumptions  imply, that the spacetime Ricci tensor
satisfies \eqref{eq:Ric_la}.

Let $\ih$ be a non-expanding null surface.  Whereas the induced
metric tensor is Lie dragged by every null vector field tangent to
$\ih$ we could see that the remaining ingredient of the geometry,
the covariant derivative, is subject to the null evolution
equation (\ref{eq:[l,D]1}, \ref{eq:[l,D]2}) implied by the
constraints. The equation depends on a choice of the null vector
field $\ell$, however, in general (and generically in the
$4$-spacetime dimensional case) a  geometry $(q_{ab},D_a)$ does
not admit any choice of $\ell$ such that $[\lie_\ell,D_a]=0$.

\begin{defn}
  An isolated null surface is a non-expanding null surface $\ih$
  equipped with a class $[\bsl]$ of tangent, null, non-vanishing vector
  fields $\bsl$ such that
  \begin{subequations}\label{eq:D_const}\begin{align}
    \lie_{\bsl} q_{ab}\ &=\ 0 \ ,  &
    [\lie_{\bsl} , D_a]\ &=\ 0 \   \tag{\ref{eq:D_const}}
  \end{align}\end{subequations}
  where $q_{ab}$ is the induced degenerate metric tensor and $D_a$ is
  the induced covariant derivative, and $\bsl,\bsl'\in[\bsl]$
  provided $\bsl'=c\bsl$ where $c$ is a constant.
\end{defn}

In this section we consider an isolated null surface
$(\ih,[\bsl])$. We assume $\ih$ is connected. Note, that given the
flow $[\bsl]$, the rotation 1-form $\w{\bsl}_a$ is defined
uniquely owing to (\ref{eq:omega'}). Obviously, it is Lie dragged
by $[\bsl]$,
\begin{equation}
  \bsl^b\lie_{\bsl} \w{\bsl}_a\ =\ \lie_{\bsl} (D_a\bsl^b)\
  =\ D_a\lie_{\bsl} \bsl^b\ =\ 0 \ .
\end{equation}
In the consequence, Theorem \ref{thm:0th} takes the familiar form
of the $0$th Law of the black hole thermodynamics,
\begin{equation}\label{eq:k_const}
  \sgr{\bsl}\ =\ \const \ ,
\end{equation}
where the value of the surface gravity depends on the choice of
$\ell\in[\ell]$ unless $\sgr{\ell}=0$.

The constraint (\ref{eq:[l,D]1}, \ref{eq:[l,D]2}) takes the following form
\begin{equation}\label{eq:[l,D]}
  D_{(a}\w{\bsl}_{c)}\ +\ \w{\bsl}_a\w{\bsl}_c
    + \frac{1}{2}\left(\Ricn{n}_{ac}-\pi^*\Ricn{n-2}_{ac}\right)\ 
  =\ 0 \ .
\end{equation}

Not surprisingly, a necessary condition is that the pullback of the space-time
Ricci tensor on $\ih$ is Lie dragged by $[\bsl]$,
\begin{equation}
  \lie_{\bsl}\Ricn{n}_{ab}\ =\ 0 \ .
\end{equation}

To understand better the meaning of the equation \eqref{eq:[l,D]}
let us apply the (local) decomposition of $D_a$ introduced in
Section \ref{sec:D_comp}. Introduce a foliation of $\ih$ preserved
by $[\bsl]$ and use the corresponding covector $n_a$, orthogonal
to the slices and normalized to an arbitrarily fixed null vector
field $\bsl^a$ generating the flow $[\bsl]$. If the derivative
$D_a$ satisfies the definition of the isolated null surface, then
the corresponding transversal expansion-shear tensor $\tS_{AB}$
defined on the slices is invariant with respect to the null flow
\begin{equation}
  \lie_{\bsl} \tS_{AB}\
  =\ \lie_{\bsl} \left( \tq^a{}_A\tq^b{}_B D_an_b \right)\
  =\ 0 \ ,
\end{equation}
because all the factors in the parenthesis are invariant. Conversely,
given a non-expanding null surface $\ih$,  a null flow $[\ell]$ generated
by nowhere vanishing vector field $\ell^a$, and one of the foliations defined
in Sec. \ref{sec:slices}, the invariance of $\w{\ell}_a$ and $\tS_{AB}$ with
respect to the null flow implies that $(\ih,[\ell])$ is an isolated
null surface.

Now, the constraint \eqref{eq:S_ev} implies
\begin{equation}\label{eq:S_const}
  \sgr{\bsl}\tS_{AB}\ =\ \tD_{(A}\tw{\bsl}_{B)}
    + \tw{\bsl}_A\tw{\bsl}_B - \fracs{1}{2}\,\Ricn{n-2}_{AB}
    + \tRicn{n}_{AB}  \ .
\end{equation}

A characterization  of the  isolated null surface depends
crucially on whether $\sgr{\bsl}$ vanishes or not, therefore we
define two types of the isolated null surfaces:
\begin{enumerate}[ (i)]
  \item {\it extremal},  if $\sgr{\bsl}=0$, or
  \item {\it non-extremal}, whenever $\sgr{\bsl}\neq0$.
\end{enumerate}
The meaning of  the constraint \eqref{eq:S_const} depends on the
type. In the non-extremal case \eqref{eq:S_const} determines
$\tS_{AB}$ given $\tq_{AB}, \tw{\bsl}_A$ and the
pullback  $\tRicn{n}_{AB}$ of the spacetime Ricci tensor
expressed by the cosmological constant and the matter energy
momentum tensor.

\begin{theorem}[Non-extremal, vacuum isolated null surface]
  Let $(\ih,\,[\bsl])$ be a non-extremal isolated null surface; suppose the
  vacuum
  Einstein equations with a cosmological constant $\Lambda$ are satisfied.
  Then, the geometry of $\ih$ is determined by the induced metric tensor
  $q_{ab}$, the rotation 1-form potential $\w{\bsl}_a$ and the
  value $\Lambda$ of the cosmological constant.
\end{theorem}

If  matter fields are present, then typically the geometry is
determined just by adding to $(q_{ab},\w{\bsl}_a, \Lambda)$ an
appropriate information on the field on $\ih$.

In the extremal case, on the other hand, equation (\ref{eq:[l,D]})
becomes a  condition on $\tw{\bsl}_{A}$, $q_{AB}$ and
$\tRicn{n}_{AB}$.

\begin{theorem}[Extremal, isolated null surface]
  Suppose $(\ih,\,[\bsl])$ is an extremal isolated null surface
  contained in $n$ dimensional spacetime;
   then, for every $n-2$-dimensional spacelike submanifold $\slc$
   transversal to the orbits of the null flow,
  the following constraint is satisfied
  \begin{equation}\label{eq:ex_constr}
    \tD_{(A}\tw{\bsl}_{B)}
      + \tw{\bsl}_A\tw{\bsl}_B - \fracs{1}{2}\,\Ricn{n-2}{}_{AB}
      + \fracs{1}{2}\tRicn{n}{}_{AB}\ 
    =\ 0 \ ,
  \end{equation}
  where $\tD_A$ and $\Ricn{n-2}{}_{AB}$ are, respectively, the
  metric, torsion free connection and the corresponding Ricci tensor
  of the metric tensor $\tq_{AB}$ induced on $\slc$.
\end{theorem}

In the vacuum case, the geometry of extremal isolated surfaces
gives rise to an equation which can be formulated in a self
contained way. Given an  $n-2$-dimensional manifold $\slc$, consider
a pair $(\tq_{AB}, \tw{\bsl}_A)$, which consists of,
respectively, a metric tensor field (of the Riemannian signature )
and a differential 1-form. The equation reads
\begin{equation}\label{eq:ex_vac}
  \tD_{(A}\tw{\bsl}_{B)}
      + \tw{\bsl}_A\tw{\bsl}_B - \fracs{1}{2}\,\Ricn{n-2}{}_{AB}\
  =\ \fracs{1}{2}\Lambda \tq_{AB} \
\end{equation}
where $\Lambda$ is the cosmological constant and $\tilde{q}_{AB}$
is still defined by (\ref{tilde}). In the case when $\slc$ is
compact and $\Lambda=0$, the equation has quite interesting
properties. They were discussed in \cite{ex} in the $n=4$ case. In
particular it was shown there that if $\slc$ is topologically a
2-sphere, then the only axially symmetric solutions are those
defined by the extremal Kerr solutions at their event horizons.
The general solution to the equation \eqref{eq:ex_vac} is not
known.

\subsection{Non-expanding null surfaces admitting a 2-dimensional
  null symmetry group} 
\label{sec:uniq_is}

A given  isolated null surface $(\ih,[\bsl])$, a priory there may
exist another null flow $[\bsl']$ defining a symmetry of the
geometry $(q_{ab},D_a)$ and being another isolated null surface
structure. In the $4$ spacetime  dimensions this non-generic case
of 2-dimensional null symmetry group was studied in detail
(see \cite{abl-g,ex}). In particular, an unexpected  relation with
the extremal isolated null surface constraints was discovered
and used in the construction of examples \cite{ex}.
 It turns out that those results can be easily generalized
to the surfaces embedded in a higher dimensional spacetime. We are
concerned with this issue in this subsection.

Suppose then, that a non-expanding null surface $\ih$ admits two
distinct isolated null surface structures   $[\bsl]$, and
$[\bsl']$. Let vector fields $\bsl$ and $\bsl'$ be generators of
the flows. There exists a real nowhere vanishing  function $f$
defined on $\ih$, such that
\begin{equation}\label{eq:f_sym}
  \bsl'\ =\ f\bsl \ .
\end{equation}
Each of the commutators $[\lie_{\bsl},D_a]$ and
$[\lie_{\bsl'},D_a]$ is represented, respectively, by the tensor
$N_{ab}$ and $N'_{ab}$ (\ref{eq:[l,D]}). According to the very
assumption made in this subsection, they both identically vanish.
On the other hand, generally one is related to another by the
following transformation law:
\begin{equation}\label{eq:N-trans}
  fN'{}_{bc} \
    = \ fN_{bc}
    + \w{\bsl}_cD_bf + \w{\bsl}_bD_cf
                       + D_bD_cf \ .
\end{equation}
If both vector fields $\bsl$ and $f\bsl$  are the symmetries of
$(q_{ab},D_a)$ then both Lie derivatives vanish. The equation
above becomes then a differential condition on the function $f$,
namely
\begin{equation}\label{eq:nuniq_trans}
  D_aD_b f + 2 \w{\bsl}_{(a} D_{b)} f\ =\ 0 \ .
\end{equation}
Contraction of the condition with $\bsl^b$ gives the equation
\begin{equation}
  D_a(\lie_{\bsl}f + \sgr{\bsl}f)\ =\ 0 \ ,
\end{equation}
which allows us to determine the null evolution of $f$
\begin{equation}
  \lie_{\bsl}f + \sgr{\bsl}f\ =\ \sgr{\bsl'}\ =\ \const  \ .
\end{equation}
By integrating this equation we obtain a solution whose form
depends on the surface gravity $\sgr{\bsl}$
\begin{equation}\label{eq:f}
  f\ =\ \begin{cases}
          B e^{-\sgr{\bsl}v} + \fracs{\sgr{\bsl'}}{\sgr{\bsl}}
            & \sgr{\bsl}\neq 0 \\
          \sgr{\bsl'} v - B  &  \sgr{\bsl}= 0 \\
        \end{cases}
\end{equation}
where $v$ is a coordinate compatible with $\bsl$ (defined via
\eqref{eq:def_v}) and $B$ is
an arbitrary real function constant along the null generators.

Note that we used the Zeroth Law according to which  the surface
gravity is constant at the surface. The Zeroth Law relies on
Stronger Energy Condition \eqref{c:energy}.

To determine the function $B$ we need to use the remaining part of
\eqref{eq:nuniq_trans}, namely its projection onto the slice
$\slc_v$ (see (\ref{eq:S_def},\ref{eq:SAB}))
\begin{equation}\begin{split}\label{eq:n_t_s}
  \tq^a{}_A\tq^b_B(D_aD_bf+2\w{\bsl}_{(a}D_{b)}f)\ 
  &=\ \tD_A\tD_Bf + 2\tw{\bsl}_{(A}\tD_{B)}f
   - \tq^a{}_A\tq^b_B D_a(\bsl^{b'}n_b+n^{b'}\bsl_b)D_{b'}f \\
  &=\ \tD_A\tD_Bf + 2\tw{\bsl}_{(A}\tD_{B)}f - \tS_{AB}\lie_{\bsl} f \ .
\end{split}\end{equation}

Without lost of the generality we can restrict ourselves to the
following cases:
\begin{enumerate}[ (i)]
  \item \label{it:nex} $\sgr{\bsl}\neq 0$
  \item \label{it:2ex} $\sgr{\bsl}=\sgr{\bsl'}=0$.
\end{enumerate}
In  both of them the equation \eqref{eq:n_t_s} is equivalent to
the following differential constraint for $B$:
\begin{equation}\label{eq:nuniq_B}
  \left[ \tD_A\tD_B + 2\tw{\bsl}_{(A}\tD_{B)} + \sgr{\bsl}\tS_{AB}
  \right] B\ =\ 0 \ .
\end{equation}

By the comparison with the constraint \eqref{eq:S_const} on the
isolated null surface geometry we can see that the term
$\sgr{\bsl}\tS_{AB}$ can be replaced by the appropriate functional
of $(\tq_{AB},\tw{\bsl_{A}},\tRicn{n}_{AB})$. The
resulting equation leads to an interesting conclusion. Since $B$
nowhere vanishes (the flows are both non-trivial and distinct) the
set of data $(\tq_{AB},\tw{\bsl}_A,\tRicn{n}_{AB},B)$
satisfies the constraint \eqref{eq:nuniq_B} if and only if the set
$(\tq_{AB},\tw{\bsl_o}_A=\tw{\bsl}_A+\tD_A\ln B,
\tRicn{n}_{AB})$ satisfies the constraint
\eqref{eq:ex_constr} for the geometry of the extremal isolated
null surface. We will go back to the consequence of this result at
the end of the the next section.

\section{Non-expanding Horizons and Isolated Horizons}
\label{sec:neh+ih} 

Thus far our considerations were purely local.
No global assumptions concerning the null surfaces topology were
made.
The specific property of a quasi-locally defined black hole is its
compact character in space-like dimensions. This notion has not
been defined on the most general level. We consider in our paper
the topologically simplest and, at the same time, the typical case
of the Cartesian product structure:

\subsection{Non-expanding horizons}
\label{sec:neh}

\begin{defn}\label{def:neh}
  A non-expanding null surface  $\ih$ in an $n$ dimensional
  spacetime $\M$ is
  called  a non-expanding horizon (NEH) if there is an embedding
  \begin{equation}
    \bas"\times\I\ \rightarrow\ \M
  \end{equation}
  such that:
  \begin{itemize}
    \item $\ih$ is the image, \item $\bas"$ is an $n-2$ dimensional
      compact manifold, \item $\I$ is the real line, \item for every
      maximal null curve in $\ih$ there is $\hat{x}\in\bas"$ such that
      the curve is the image of $\{\hat{x}\}\times \I$.
  \end{itemize}
\end{defn}
The {\it base space} $\bas$ defined as the space of all the maximal null
curves in $\ih$ can be identified with the manifold $\bas"$ given
an embedding used in Definition \ref{def:neh}. Whereas the
embedding is not unique, the manifold structure defined in this
way  on $\bas$ is unique. There is also a uniquely defined
projection
\begin{equation}
  \pi: \ih\ \rightarrow\ \bas \ .
\end{equation}

In this section we consider a NEH $\ih$. Of course it inherits all
the properties of the non-expanding null surfaces. The following
theorems are applications of the results of Section \ref{sec:nes}
to the non-expanding horizons.

The first theorem summarizes the properties following from the
weaker energy assumption \eqref{eq:Tll>0}:

\begin{theorem}\label{thm:nehtll} 
  Suppose  $\ih$ is a non-expanding horizon
  in a spacetime $\M$. Suppose at $\ih$ the spacetime  Einstein
  field equations hold and the matter  fields satisfy the condition
  (\ref{eq:Tll>0}). Let  $\ell$ be an arbitrary null vector field
  tangent to $\ih$ (in items (iii-vii) below); then:
  \begin{enumerate}[ (i)]
    \item there is a metric tensor field $\hq_{AB}$ (called projective)
      defined on the base space $\bas$, such that the degenerate
      metric tensor $q_{ab}$ induced in $\ih$ by the space-time metric
      tensor is given by the pullback,
      \begin{equation}\label{eq:hq}
        q_{ab}\ =\ \pi^{*}\hq_{ab} \ ;
      \end{equation}
    \item there is a covariant derivative $D_a$ defined in the tangent
      bundle $T\ih$ such that, for every two vector fields $X,Y$,
      \begin{equation}
        D_XY\ =\ \nabla_XY \ ,
      \end{equation}
      where $\nabla_\alpha$ is the space-time covariant derivative;
    \item there is a 1-form $\w{\ell}_a$ (called the rotation 1-form
      potential) defined on $\ih$ such that
      \begin{equation}
        D_a\ell^b\ =\ \w{\ell}_a\ell^b \ ;
      \end{equation}
    \item the rotation 2-form (invariant)
      \begin{equation}
        \Omega_{ab}\ :=\  D_a\w{\ell}{}_b - D_b\w{\ell}{}_a
      \end{equation}
      is uniquely is independent of $\ell$;
    \item The rotation 1-form potential and the self acceleration
      $\sgr{\ell} := \ell^a\w{\ell}_a$ of $\ell$ satisfy
      \begin{equation}\label{eq:pre0thH}
        \lie_{\ell}\w{\ell}_a\ =\ D_a\sgr{\ell} + \Ricn{n}_{ab}\ell^b\
      \end{equation}
    \item the infinitesimal Lie transport of $D_a$ with respect to the null
      flow of $\ell$ is the following tensor:
      \begin{equation}\label{[l,D]ih}
        [\lie_\ell,D_a]^b_c\ =\ \ell^b\left(
        D_{(a}\w{\ell}{}_{c)}\ +\ \w{\ell}{}_a\w{\ell}{}_c
        + \frac{1}{2}\left(\Ricn{n}_{ac}-\pi^*\Ricn{n-2}_{ac}\right)\right)
      \end{equation}
      where $\Ricn{n-2}_{AB}$ is the Ricci tensor of the metric tensor
      $\hq_{AB}$ 
    \item \label{it:Ric1} the following components of the pullback
      onto $\ih$ of the  spacetime Ricci and Riemann tensor vanish:
      \begin{equation} \label{it:R_tang}
        \ell^\alpha\ell^\beta\Ricn{n}_{\alpha\beta}\ =\ 0\ 
        =\ \ell^\alpha\Riem{n}_{\alpha bcd} \ .
      \end{equation}
  \end{enumerate}
\end{theorem}

In the previous sections we also considered Stronger Energy
Condition \ref{c:energy}. The following theorem summarizes its
consequences for a non-expanding horizon:

\begin{theorem}\label{thm:nehtltl}
  Suppose all the assumptions of Theorem \ref{thm:nehtll} are
  satisfied and additionally the matter fields at $\ih$ satisfy
  Stronger  Energy Condition \ref{c:energy}; then:
  \begin{itemize}
    \item On the base space $\bas$, there is a uniquely defined 2-form
      $\hat{\Omega}_{AB}$ such that the rotation 2-form invariant is its
      pullback
      \begin{equation}\label{eq:OmegaPull}
        \Omega_{ab}\ =\ \pi^*\hat{\Omega}_{ab} \ ;
      \end{equation}
    \item the rotation 1-form potential and the self
      acceleration satisfy
      \begin{equation}\label{eq:0thH}
        \lie_{\ell}\w{\ell}{}_a\ =\ D_a\sgr{\ell} \ ;
      \end{equation}
    \item the pullback of the space-time Ricci tensor and,
      respectively, the space-time Weyl tensor onto $\ih$ is transversal
      to the null direction tangent to $\ih$,
      \begin{equation}
        \ell^\alpha\Ricn{n}_{\alpha b}\ =\ 0\ 
        =\ \ell^\alpha\Weyl{n}_{\alpha bcd} \ .
      \end{equation}
  \end{itemize}
\end{theorem}
We will consider now the non-expanding horizons such
Theorem \ref{thm:nehtltl} applies.

In the case of the non-expanding horizons, there are globally
defined, nowhere vanishing null vector fields $\ell^a$ tangent to
$\ih$ at our disposal. In particular, there is a vector field
$\ell_o^a$ of the identically vanishing self acceleration,
$\sgr{\ell_o}=0$. There is also a null vector field $\ell^a$ of
$\sgr{\ell}$ being an arbitrary constant,\footnote{ The first one,
  $\ell_o$  can be defined by fixing appropriately affine parameter
  $v$ at each null curve in $\ih$. Then, the second vector field is
  just $\ell=v\ell_o$.}
\begin{equation}\label{eq:kappaconst}
  \sgr{\ell}\ =\ {\rm const} \ .
\end{equation}
The vector field $\ell^a$ can vanish  in a harmless, for our
purposes, way on an $n-2$ dimensional section of $\ih$ only. We
fix one of the vector fields $\ell^a$ (including ($\ell_o^a$))
throughout this section. We will also use a coordinate $v$
compatible with the vector field $\ell^a$ ( $\ell^aD_av=1$ ), and
the covector field $n_a$ ( $=-D_av$ ), both introduced in Section
\ref{sec:slices} defined on $\ih$ (except the zero slice of
$\ell$). It follows from the Zeroth Law (\ref{eq:0thH}) that the
rotation 1-form potential is Lie dragged by $\ell$,
\begin{equation}
  \lie_\ell \w{\ell}_a\ =\ 0 \ .
\end{equation}
We discuss below two independent consequences of this fact.   The
first one is the existence of a new invariant of the geometry of
$\ih$, a certain harmonic 1-form invariantly defined on the base
manifold $\bas$. The second one concerns the degrees of freedom of
a general vacuum solution $(q_{ab},D_a)$ of the constraints
(\ref{eq:nes_q}, \ref{eq:[l,D]1}, \ref{eq:[l,D]2}).

\subsubsection{Harmonic invariant}
\label{harmonic}

It turns out, that $\w{\ell}_a$
defines on the base space $\bas$ a unique  1-form depending only
on the geometry $(q_{ab},D_a)$ of $\ih$. Indeed, given the
function $v$, there is a differential 1-form field
$\hw{\ell}_A$ defined on $\bas$ and called the
projective rotation 1-form potential, such that
\begin{equation}\label{eq:omega}
  \w{\ell}_a\ =\ \pi^*\hw{\ell}_a\
    +\ \sgr{\ell}D_av \ .
\end{equation}
The 1-form $\hw{\ell}_A$ is not uniquely defined,
though. It depends on the choice of the function $v$ compatible
with $\ell^a$, and on the choice of $\ell^a$ itself. Given
$\ell^a$, the freedom is in the transformations
\begin{subequations}\begin{align}
  v\ &=\ v' + B,\quad \lie_\ell B \ =\ 0\label{eq:v'} \ , \\
  \hw{\ell}'_A \ &=\  \hw{\ell}_A+\hD_A B \ .\label{eq:tomega'}
\end{align}\end{subequations}
The transformations $\ell'^a=f\ell^a$ which preserve the condition
(\ref{eq:kappaconst}) are given by \eqref{eq:f}, and it can be
shown using (\ref{eq:omega'}), that  the only possible form of the
corresponding $\tw{\ell'}_A$ is again that of (\ref{eq:tomega'})
with possibly different function $B$. Therefore, if we apply the
(unique) Hodge decomposition onto the exact, the co-exact, and the
harmonic part, respectively, to $\hw{\ell}_A$,
\begin{equation}
  \hw{\ell}{}_A\ =\ \hw{\ell}{}_A^{\rm ex}
  + \hw{\ell}{}_A^{\rm co} + \hw{\ell}{}_A^{\rm ha}\ ,
\end{equation}
then the parts $\hw{\ell}{}_A^{\rm co}$ and
$\hw{\ell}{}_A^{\rm ha}$ are invariant, that is
determined by the geometry $(q_{ab},D_a)$ of $\ih$ only. The
co-exact part is determined by the already defined invariant
2-form (\ref{eq:Omega}), via
\begin{equation}
  \hat{\Omega}_{AB}\ =\ \hD_A \hw{\ell}{}_B^{\rm co}
  - \hD_B \hw{\ell}{}_A^{\rm co}\ .
\end{equation}
The harmonic part of  $\hw{\ell}{}_A$ is the new invariant. It did
not occur  in  the case of spherical $\bas$ considered in
\cite{abl-g}. In the case of a general topology of $\bas$, the
invariant may be relevant. There are known non-trivial examples of
black holes with non-spherical base spaces. For instance, in
5-dimensions there exists asymptotically flat, regular,
axi-symmetric solutions (see \cite{ring} for details) of the
horizon base space topology $S^1\times S^2$. The space of harmonic
1-forms is finite-dimensional, so the degrees of freedom
identified with the harmonic component of the rotation 1-form
potential are global in the character.

\subsubsection{Degrees of freedom}
\label{degrees}

Let  $\ell^a$, $v$ and $n_a$ be still the same, respectively,
vector field, a compatible coordinate and a covector field
introduced beneath Theorem \ref{thm:nehtltl}. The covariant
derivative $D_a$ is characterized by the elements defined in
Section \ref{sec:D_comp}, subject to the constraints
(\ref{eq:[l,D]1}, \ref{eq:[l,D]2}). Suppose the vacuum Einstein
equations with a (possibly zero) cosmological constant are
satisfied on $\ih$.  \\
{\it The geometry $(q_{ab},D_a)$ can be completely
characterized by the following data:
\begin{enumerate}[ (i)]
  \item defined on the space of the null geodesics $\bas$:
     \begin{itemize}
       \item the projective metric tensor $\hq$ (\ref{eq:hq})
       \item the projective rotation 1-form potential
             $\hw{\ell}{}_A$
             (\ref{eq:omega})
       \item the projective transversal expansion-shear data $\hS^o_{AB}$
             (see (\ref{eq:transexpsh}) below)
     \end{itemize}
   \item the values of the surface gravity $\sgr{\ell}$ and the
         cosmological constant $\Lambda$,
\end{enumerate}}
where the projective transversal expansion-shear data
$\hS^o_{AB}$ is a tensor defined on $\bas$  by the following form
of a general solution of \eqref{eq:S_ev},
\begin{equation}\label{eq:transexpsh}
  \tS_{AB}\ =\ \begin{cases}
    \left(\hD_{(A}\hw{\ell}_{B)}
      + \hw{\ell}_A\hw{\ell}_B - \fracs{1}{2}\,\Ricn{n-2}{}_{AB}
      - \fracs{1}{2}\Lambda \hq_{AB}\right)v + \pi^*\hS^o{}_{AB} &
      \text{for } \ \ \sgr{\ell}=0 \ , \\
    e^{-\sgr{\ell} v}\pi^*\hS^o{}_{AB}
      + \frac{1}{\sgr{\ell}}\left(\hD_{(A}\hw{\ell}_{B)}
      + \hw{\ell}_A\hw{\ell}_B - \fracs{1}{2}\,\Ricn{n-2}{}_{AB}
      - \fracs{1}{2}\Lambda \hq_{AB}\right)& \text{otherwise.}\\
  \end{cases}
\end{equation}
A part of data depends on the choice of the vector field $\ell^a$
and the compatible coordinate $v$. Given $\ell^a$ such that
$\sgr{\ell}\not=0$, the compatible coordinate $v$ can be fixed up
to a constant by requiring that the exact part in the Hodge
decomposition of the projective rotation 1-form potential
$\hw{\ell}{}_A$ vanishes (see Section \ref{harmonic}). The vector
$\ell^a$ itself, generically, can be fixed up to a constant factor
by requiring that the projective transversal expansion-shear data
$\hS^o{}_{AB}$ be traceless. Indeed,  the transformations of
$\ell^a$ such that $\sgr{\ell}$ remains a non-zero constant are
given by  (\ref{eq:f_sym}, \ref{eq:f}). They are accompanied by
the transformations
$\lie_{\bsl}S_{ab}\mapsto\lie_{\bsl'}S'{}_{ab}$ determined by
\eqref{eq:N-trans}. Contraction of the mentioned equation with the
tensor $\tq^{ab}$ defined as follows
\begin{subequations}\label{eq:tq-a}\begin{align}
  \pi_{*}\tq^{ab} &= \hq^{ab} \ , &
  \tq^{ab}n_b &= 0 \ , \tag{\ref{eq:tq-a}}
\end{align}\end{subequations}
and the assumption that $\tq^{ab}\lie_{\ell'}S_{ab}=0$ (equivalent
to $\hq^{AB}\hS^o{}_{AB}=0$) produces the following gauge
condition defined on the slices:
\begin{equation}\label{eq:trN-trans}
  \left[ \tD^2 + 2\tw{\bsl}^A\tD_A
       + \tq^{AB}(\sgr{\bsl}\tS_{AB} + \lie_{\bsl}\tS_{AB}) \right] B\ 
   =\ \fracs{\sgr{\bsl'}}{\sgr{\bsl}}
     e^{\sgr{\bsl}v}\tq^{AB}\lie_{\bsl}\tS_{AB}
\end{equation}
According to the Zeroth Law and \eqref{eq:transexpsh} the above
equation defines at each slice the same constraint for a NEH's
geometry, and can be rewritten in terms of the objects defined on
the base space $\bas$. Hence, the condition that $\hS_{AB}$ be
traceless takes the form of the following elliptic equation on the
function $B$,
\begin{equation}\label{trS0-trans}
  \left[ \hD^2 + 2\hw{\bsl}^A\hD_A
       + \text{div}\hw{\bsl} + |\hw{\bsl}|^2
       - \fracs{1}{2}\hq^{AB}\Ricn{n-2}_{AB} - \fracs{n-2}{2}\Lambda
  \right] B\ =\ \sgr{\bsl'}\hq^{AB}\hS^o{}_{AB}
\end{equation}
where $\text{div}\hw{\bsl}:=\hD^A\hw{\bsl}_A$ and
$|\hw{\bsl}|^2:=\hw{\bsl}^A\hw{\bsl}_A$. The equation
generically has a unique solution. Finally, the remaining
re-scaling freedom by a constant can be removed by fixing the
value of the surface gravity $\sgr{\ell}$ arbitrarily (the area of
$\ih$ can be used as a quantity providing the appropriate units).

\subsubsection{Abstract non-expanding null surface / horizon
  geometry}
\label{sec:abstract}

Non-expanding null surface  geometry can be  defined abstractly.
Consider an $n-1$-dimensional  manifold $\ih$. Let $q_{ab}$ be a
symmetric  tensor of the signature $(0,+...+)$. Let $D_a$ be a
covariant, torsion free derivative such that
\begin{equation}
  D_aq_{bc}\ =\ 0 \ .
\end{equation}
A vector  $\ell^a$ tangent to $\ih$ is called null whenever
\begin{equation}
  \ell^aq_{ab}\ =\ 0 \ .
\end{equation}
Even though we are not assuming any  symmetry, every null vector
field $\ell^a$ is a symmetry of $q_{ab}$:
\begin{lem}
  Suppose $\ell^a$ is a null vector field tangent to $\ih$; then
  \begin{equation}
    \lie_\ell q_{ab}\ =\ 0 \ .
  \end{equation}
\end{lem}
Despite of the fact, that Lemma is quite surprising, the proof is
not difficult. We leave it to the interested reader.

Given a null vector field $\ell^a$, we can repeat the definitions
of Section \ref{sec:nes}, and associate to it the surface gravity
$\sgr{\ell}$, and the rotation 1-form potential $\w{\ell}$. Now, a
vacuum Einstein constraint can be defined as an equation on the
geometry  $(q_{ab},D_a)$ per analogy with the non-expanding null
surface case. To spell it out we need one more definition.
Introduce on $\ih$ a symmetric tensor $\Ricn{n-2}_{ab}$, such that
for every $n-2$-subsurface contained in $\ih$ the pullback of
$\Ricn{n-2}_{ab}$ to the subsurface coincides with the Ricci tensor
of the induced metric, provided the pullback $\tq_{AB}$ of
$q_{ab}$ is a non-degenerate metric tensor.  The vacuum constraint
is defined as
\begin{equation}\label{eq:einsteinvac}
  [\lie_\ell,D_a]_c^b\ =\ \left(D_{(a}\w{\ell}{}_{c)}\ +\
  \w{\ell}{}_a\w{\ell}{}_c\ -\ \fracs{1}{2}\Lambda q_{ac}\right) \ -\
  \fracs{1}{2}\Ricn{n-2}_{ac} \ ,
\end{equation}
and it involves an arbitrary  cosmological constant $\Lambda$.

Suppose now, that
\begin{equation}
  \ih\ =\ \bas\times \mathbb{R}\ ,
\end{equation}
and the tensor $q_{ab}$ is the product tensor defined naturally by
a metric tensor $\hq_{AB}$ defined in $\bas$ and the
identically zero tensor defined in $\mathbb{R}$. The analysis of
Sections \ref{harmonic} and \ref{degrees} can be repeated to
solutions of the vacuum Einstein constraint
(\ref{eq:einsteinvac}). Again the base space $\bas$ is equipped
with the data of in Section \ref{degrees}, that is the projective:
metric tensor $\hq_{AB}$, rotation 1-form potential
$\hw{\ell}_{A}$,  transversal expansion-shear data $\hS^o{}_{AB}$.
Completed by the values of the surface gravity $\sgr{\ell}$ and
the cosmological constant $\Lambda$ the data is free, in the sense
that every data set defines a single solution $(q_{ab},D_a)$.

\subsection{Isolated Horizons}
\label{sec:ih}

\begin{defn}\label{def:ih}
  An isolated null surface $(\ih,[\bsl])$ such that the surface $\ih$ is a
  non-expanding horizon  is called an isolated horizon (IH).
\end{defn}

Consider an arbitrary isolated horizon $(\ih,[\bsl])$. A generator
$\bsl$ of the null symmetry is defined globally on $\ih$, and it
is unique modulo the re-scaling $\bsl'=a_0\bsl$ by a constant
$a_0$. Therefore, the rotation 1-form potential $\wI$ is defined
globally on $\ih$ and in an independent of the choice of
$\bsl'\in[\bsl]$ manner (hence we will drop in this section the
suffix $(\bsl)$ at $\wI$ but keep it at the surface gravity). It
follows from Section \ref{sec:iso} that for every isolated horizon
$(\ih,[\bsl])$ the rotation 1-form potential is Lie dragged by the
vector field $\bsl$,   additionally the energy condition
(\ref{eq:Tll>0}) is satisfied necessarily and the left hand side
of (\ref{[l,D]ih}) is assumed to be zero. In conclusion:

\begin{enumerate}[ (i)]
  \item The rotation 1-form potential is Lie-dragged by the horizon symmetry
        $\bsl$
        \begin{equation}\label{eq:lie_wH}
          \lie_{\bsl}\wI_a\ =\ 0 \ ,
        \end{equation}
  \item if the matter fields  satisfy Stronger Energy Condition
        \ref{c:energy} on $\ih$, then the surface gravity
        $\sgr{\bsl}$ is
           constant,
           \begin{equation}
           \sgr{\bsl}\ =\ \const\ .
           \end{equation}
  \item the pullback of the space-time
        Ricci tensor on $\ih$ is Lie dragged by $[\bsl]$,
        \begin{equation}\label{eq:lie_R}
          \lie_{\bsl}\Ricn{n}_{ab}\ =\ 0 \ ,
        \end{equation}
  \item in the case when Stronger  Energy Condition \eqref{c:energy} holds,
        the tensor $S_{ab}$ has the following
        form
        \begin{equation}
          S_{ab}\ =\ \pi^{*}\hS_{ab} - 2\wI_{(a}n_{b)} -
          \sgr{\bsl}n_an_b \ ,
        \end{equation}
         where $\hS_{AB}$ is a symmetric tensor defined on
         $\bas$;
  \item The constraint \eqref{eq:S_ev} applied to $S_{ab}$ above reads
        \begin{equation}\label{eq:S_constH}
          \sgr{\bsl}\hS_{AB}\ =\ \hD_{(A}\hwI{}_{B)}
          + \hwI_A\hwI_B - \fracs{1}{2}\,\Ricn{n-2}{}_{AB}
          + \frac{1}{2}\hRicn{n}{}_{AB}  \ ,
        \end{equation}
        where by $\hRicn{n}{}_{AB}$ we denoted the tensor uniquely
        defined  on $\bas$ such that
        $\pi^*\hRicn{n}{}_{ab}=\Ricn{n}{}_{ab}$.
\end{enumerate}

The classification of
the isolated null surfaces with respect to whether $\sgr{\bsl}$
vanishes or not applies to the isolated horizons, therefore we
call an isolated horizon extremal whenever $\sgr{\bsl}=0$, and
non-extremal otherwise.

\subsubsection{Degrees of freedom: the non-extremal case}

Suppose the vacuum Einstein equations (with a possibly non-zero
cosmological constant) hold on an isolated horizon $(\ih,[\bsl])$,
and
\begin{equation}
  \sgr{\bsl}\ \not=\ 0 \ .
\end{equation}

 Since $\ih$ is a non-expanding horizon, its geometry can be
characterized by the data $(i)$ and $(ii)$ discussed in Section
\ref{degrees}. Now, however, the data satisfies an extra
constraint following from (\ref{eq:S_const}), namely
\begin{equation}
  \hS^o_{AB} = 0
\end{equation}
in (\ref{eq:transexpsh}). Therefore, in the non-extremal isolated
horizon case, given a generator $\bsl$ of the flow $[\bsl]$, the
geometry $(q_{ab},D_a)$ is completely determined by the projective
metric and the projective 1-form potential
$(\hq_{AB},\hwI_{A})$ defined on the base manifold
$\bas$, provided the surface gravity $\sgr{\bsl}$ and the
cosmological constant are given. The discussion of the gauge
degrees of freedom in the data of Section \ref{degrees} applies,
except, that in this case the null flow $[\bsl]$ is given. The
data $\hq_{ab}$ and  $\hwI_A$ is free in the sense of
Section \ref{sec:abstract}.

\subsubsection{Degrees of freedom - the extremal case}

In the extremal case, on the other hand, the vacuum isolated
horizon constraints \eqref{eq:S_const} do not constrain the
projective transversal expansion-shear data $\hS^o_{AB}$ at all. On
the other hand, the projective metric tensor $\hq_{AB}$ and the
projective rotation 1-form potential $\hwI_A$ necessarily satisfy
a constraint
\begin{equation}\label{eq:ex_constrH}
\left(\hD_{(A}\hwI_{B)}
    + \hwI_A\hwI_B - \fracs{1}{2}\,\Ricn{n-2}{}_{AB}
    - \fracs{1}{2}\Lambda \hq_{AB}\right) \ =\ 0\ .
\end{equation}
The general solution to this equation is not known even in the
case of 4-dimensional spacetimes, however the number
$\fracs{1}{2}n(n-1)$ of the equations
 is in the space of the solutions to the
constraints \eqref{eq:D_const} equal to the number of the
independent variables: $\fracs{1}{2}n(n-1)$ for the metric
\cite{metr_var} plus $n-2$ for the rotation. Therefore, one can
expect, that extremal isolated horizons should be described by the
global degrees of freedom.

In the extremal case, as opposed to the non-extremal case, the
projective rotation 1-form potential $\hwI_A$ is uniquely
defined on $\bas$, including the exact part. The projective
transversal expansion-shear  tensor $\hS_{AB}$, on the other
hand, is still the choice of the compatible coordinate $v$ dependent.
The transformation of projective tensor $\hS_{AB}=\hS^o{}_{AB}$ is
\begin{equation}\label{eq:S_sl_trans}
  \hS_{AB}\ \to\ \hS_{AB}
    + \left[ \hD_{A}\hD_{B} +
    2\hwI_{(A}\hD_{B)} \right] f \ .
\end{equation}
The trace of this equation with respect to $\hq_{AB}$ becomes
an elliptic PDE for the function $f$. Therefore generically there is a
possibility 
to distinguish the coordinate $v$ (and the corresponding family of
sections of $\ih$) by the requirement that the trace of
$\hS_{AB}$, as well as the trace of $\tS_{AB}$ be zero.

Finally, in the sense of Section \ref{sec:abstract}, the degrees
of freedom in the space of the extremal isolated horizons,
solutions to the constraints \eqref{eq:D_const}) are given by the
solutions $(\hq_{AB},\hwI_A)$ of the
constraint \eqref{eq:ex_constrH}, and the traceless part of the
transversal expansion shear tensor $\hS_{AB}$.

\subsection{Non-expanding horizons with  2-dimensional
  null symmetry group}

In four dimensional case (see \cite{abl-g}) there exist
non-expanding horizons which admit 2-dimensional group of the null
symmetries.  In Section \ref{sec:uniq_is} we investigated the
conditions for the existence of more than one null symmetry of an
isolated null surface in arbitrary dimension. In this section we
will investigate further the geometries of the isolated horizons
admitting more than one null symmetry. We will show the following
theorem:
\begin{theorem} Suppose $\ih$ is a non-expanding horizon
  which admits two distinct isolated horizon structures. Suppose the
  energy condition \eqref{c:energy} holds on $\ih$. Then, $\ih$
  admits an extremal isolated horizon structure $[\bsl']$ and a
  compatible coordinate $v'$ such that the corresponding transversal
  expansion-shear tensor $\tS_{AB}$ identically vanishes at $\ih$.
\end{theorem}
\begin{proof}
Let $[\bsl]$ and $[\bsl']$ be two different isolated horizon
structures at $\ih$ generated by  $\bsl$ and $\bsl'$ respectively.
According to the Zeroth Law, the surface gravities $\sgr{\bsl}$
and $\sgr{\bsl'}$ are constant on the horizon. Suppose
\begin{equation}
  \sgr{\bsl}\ \not=\ 0\ .
\end{equation}
Let $v:\ih\rightarrow\mathbb{R}$ be a compatible coordinate.
The relation between $\bsl$ and $\bsl'$ is
\begin{equation}\label{eq:f_symH}
  \bsl'\ =\ f\bsl \ .
\end{equation}
where the function $f$ is of the  form:
\begin{equation}\label{eq:fH}
  f\ =\ \begin{cases}
          B e^{-\sgr{\bsl}v} + \fracs{\sgr{\bsl'}}{\sgr{\bsl}}
            & \sgr{\bsl} \neq 0 \\
          \sgr{\bsl'} v - B  &  \sgr{\bsl}= 0, \\
        \end{cases}
\end{equation}
where the function $B$ is constant along the null curves in $\ih$,
and  the necessary and sufficient  condition (\ref{eq:nuniq_B})
for the function $B$ thought of as a function $B:\bas\rightarrow
\mathbb{R}$, can be rewritten in terms of the data defined on the
base manifold $\bas$ of the null curves:
\begin{equation}\label{eq:nuniq_BH}
  \left[ \hD_A\hD_B + 2\hw{\bsl}_{(A}\hD_{B)}
  + \hD_{(A}\hw{\bsl}_{B)}
    + \hw{\bsl}_A\hw{\bsl}_B -
    \fracs{1}{2}\,\Ricn{n-2}{}_{AB}
    + \fracs{1}{2}\hRicn{n}{}_{AB} \  \right] B\ 
   =\ 0 \ .
\end{equation}
where $\hD_{(A}\hw{\bsl}_{B)}$ should be considered as a tensor not
an operator.

Note, however, the function $B$ is independent of the surface
gravity of another vector field we construct with, therefore,
either $\sgr{\bsl''}=0$, or, given the functions $f''$ and $B$, we
can define a new function $f'$,
\begin{equation}\label{eq:nex2ex}
  f'\ :=\ B e^{-\sgr{\bsl} v}\ .
\end{equation}
Then, the vector field $\bsl'=f'\bsl$ defines an extremal isolated
horizon.  We will show now, that there is a coordinate
$v':\ih\rightarrow\mathbb{R}$ compatible with $\bsl'$ such that
the corresponding projective transversal expansion-shear data
$\hS'^o_{AB}$ is identically zero.
According to the equation
\eqref{eq:nex2ex} the general form of a coordinate $v'$ compatible
with $\bsl'$ is
\begin{equation}\label{eq:v2v}
  v'\ =\ (\sgr{\bsl}B)^{-1}( e^{\sgr{\bsl}v} -1 ) + v'_0
\end{equation}
where $v'_0$ is a function constant along the null curves in
$\ih$. Let us choose $v'_0$ to be
\begin{equation}
  v'_0\ :=\ (\sgr{\bsl}B)^{-1} \ .
\end{equation}
Then the correspondence between the vector fields $n'_a=-
D_av'$ and
$n_a=-D_a v$ can be described by the following equation:
\begin{equation}\label{eq:n2n}
  n'_a\ =\ v'(\sgr{\bsl}n_a-D_a\ln B) \ .
\end{equation}
The covariant derivative of the above equation
determines the transformation $S_{ab}\mapsto S'{}_{ab}$:
\begin{equation}
  \frac{1}{v'}S'{}_{ab} - \frac{1}{{v'}^2}n'{}_an'{}_b\
  =\ S_{ab} - D_aD_b\ln B \ .
\end{equation}
Taking the lie derivative of this formula with respect to $\bsl$
and taking into account that
$\lie_{\bsl'}S'{}_{ab}=\lie_{\bsl}S_{ab}=0$ we get the following
result:
\begin{equation}
  S'{}_{ab}\ =\ \w{\bsl'}_{(a}n'{}_{b)} \ .
\end{equation}
Therefore the pullback $\tS'_{AB}$  of $S'{}_{ab}$ identically
vanishes, and so does the projective part $\hS'_{AB}$ defined
by (\ref{eq:transexpsh}).
\end{proof}

In the case when both the symmetries $\bsl$ and $\bsl'$ admits
extremal isolated horizon structures the equation \eqref{eq:nuniq_BH}
takes the following form:
\begin{equation}\label{eq:nuniq_BH_ex}
  \left[ \hD_A\hD_B + 2\hw{\bsl}_{(A}\hD_{B)} \right] B\ =\ 0 \ .
\end{equation}
Together with the constraint \eqref{eq:ex_constrH} the equation above
forms an overdefined system involving the data $\hq,\hw{\bsl}$.
The non-existence of the solutions to the system in the case of the
horizon embedded in an 4-dimensional electrovac spacetime with
vanishing cosmological constant was shown in \cite{abl-g,ex},
however one can not expect to repeat this result in the general case.
It seams that the answer for the question whether the solutions to the
system (\ref{eq:ex_constrH},\ref{eq:nuniq_BH_ex}) do exist
require an analysis for each case of the assumed dimension and the topology
of a horizon base space separately.

\section{Conclusion}

It turns out that the basic properties of  null, non-expanding
surfaces are not sensitive on the space-time dimension. We have
discussed only those properties which were found relevant in the
$4$ dimensional case. The exception is the characterization of the
surfaces admitting a 2-dimensional group of null symmetries and
the relation with the extremal isolated surface constraint.

A new element  in the characterization of the non-expanding, null
surfaces is is the harmonic 1-form invariant defined by the
rotation 1-form potential on the space of the null generators of
the surface.

\section{Acknowledgments}

We would like to thank Abhay Ashtekar, Badri Krishnan, Piotr
Chrusciel, Jose Jaramillo, Jacek Jezierski and Vojtech Pravda for
discussions. This work was supported in part by the Polish
Committee for Scientific Research (KBN) under grants no: \mbox{2
P03B 127 24}, \mbox{2 P03B 130 24}, the National Science
Foundation under grant  0090091, and the Albert Einstein Institute
of the Max Planck Society.

\appendix

\section{Remarks on the exotic matter case}

In the development of the objects describing the geometrical structure of
the non-expanding and isolated horizons there was assumed, that the energy
condition \ref{c:energy} holds for the matter fields on the horizon. From
the other hand one may need to deal with the models in which considered 
condition has to be dropped. Then the
question arises: how many of the structures developed here still apply ?
The current section is an attempt to answer this question.

\subsection{Non-expanding horizons}

In this paper the energy condition \eqref{c:energy} was in fact used
to develop the identity \eqref{eq:Ric_la} only. As the mentioned identity
is equivalent to the following condition
\begin{equation}\label{eq:T_al}
  T_{ab}\ell^b = 0 \ ,
\end{equation}
involving the pull-back of $T_{\mu\nu}$ onto the horizon, even if the
condition \ref{c:energy} is not fulfilled, all the statements still
apply as long as the equation \eqref{eq:T_al} holds. As a
'toy'\footnote{This particular
  case is already exhibited in this paper as the Einstein equation with
  cosmological constant was considered. We use it only to illustrate existence
  of possible models for which \eqref{eq:T_al} holds and
  Condition \ref{c:energy} doesn't. The real applications would be the
  'varying cosmological constant' models.}
example of the matter field satisfying \eqref{eq:T_al} we can consider the
cosmological constant represented as a matter field
$T_{\mu\nu}=-\Lambda g_{\mu\nu}$. For negative $\Lambda$ the Stronger Energy
Condition \ref{c:energy} doesn't hold any longer ($-T^{\nu}{}_{\nu}\ell^{\nu}$
is past-oriented) but the equality \eqref{eq:T_al} is still true.

Consider now the most general situation when nothing about the
energy-momentum tensor is assumed. Then the statements of the Theorem
\ref{thm:nehtll} are no longer true:
The shear of the NEH doesn't have to vanish because the
$|\sgv{\ell}|^2$ term in the Raychaudhuri equation can be balanced by
the negative $\Ricn{n}{}_{\ell\ell}$. The presence of shear affects
entire geometrical structure. For example only the area form of the
metric tensor $\bar{q}_{AB}$ is preserved by the null flow
\begin{equation}
  \lie_{\ell}\bar{\epsilon}\ =\ 0 \ .
\end{equation}
Moreover the spacetime covariant derivative does not preserve the tangent
bundle of the horizon. Therefore the internal connection $D$ of the horizon
must be introduced the other way than \eqref{eq:D_def} and none of the
statements in Section \ref{sec:neh} will hold.

One of methods to deal with the problem is to restrict investigated
objects to the {\it non-expanding shear-free horizons} defined as the null
surfaces equipped with a metric $q$ preserved by the null flow. The
restriction makes sense as the NEH's admitting the isolated horizon
structure (which includes also Killing horizons) necessarily
have to belong to this class. Note that for that class of the horizons
the condition \eqref{eq:Tll>0} is satisfied due to Raychaudhuri
equation so the discussion in the main body of the paper apply here.

The other restriction we can make is to impose the weaker energy condition
\eqref{eq:Tll>0} only. That case has been investigated thorough the main
part of the paper: The statements of the Theorem \ref{thm:nehtll}
are true for the considered NEH, whether the ones for the Theorem
\ref{thm:nehtltl} are not. Note that however the horizon can be no longer
of the type $II$ in principal classification, the component
$\Weyl{n}_{a\ell b\ell}$ still vanishes, so the horizon is at least
of the type $I$ (remaining algebraically special).

\subsection{Isolated horizons}

If we assume that the horizon admits an isolated horizon structure
its shear and Ricci component $\Ricn{n}{}_{\ell\ell}$ vanish due to
existence of a null symmetry without any assumption imposed on the
energy-momentum tensor of the matter fields.

Because of the modification of the Zeroth Law the statement
$\sgr{\bsl}=\const$ must be replaced by the following one:
\begin{equation}
  \partial_a\sgr{\bsl}\ =\ - \Ricn{n}{}_{ab}\bsl^b \ ,
\end{equation}
so the 'surface gravity' defined as $\bsl^a\wI_a$ becomes a function
constant along the null generators
\begin{equation}\label{eq:lie-sgr}
  \lie_{\bsl}\sgr{\bsl}\ =\ 0 \ .
\end{equation}
Now without any additional energy assumptions the division on the
non-extremal and extremal IH structures is no longer valid as the
structures with $\sgr{\bsl}=0$ at some open subset of $\slc$ and
$\sgr{\bsl} \neq 0$ elsewhere are possible. Therefore the structure
of the constraint (\ref{eq:S_constH}) (so the structure of the local
degrees of freedom) can be different at distinct open subsets of the 
horizon base space.

The problem of classification (and description of the degrees of freedom)
can be dealt of by imposing other assumptions which are satisfied by some
class of an exotic matter fields.

As the energy-momentum tensor of the matter fields is necessarily
divergence-free the following constraint is true on the horizon
\begin{equation}\label{c:div}
  \bsl^{\mu}T_{\mu\nu}{}^{\nu}\ =\ 0 \ .
\end{equation}
Using the identity \eqref{eq:lie_R} we can after simple calculations
express the condition above as an $n-2$ dimensional differential
equation defined on the slices of the horizon foliation preserved by a
flow $[\bsl]$:
\begin{equation}\label{eq:PDE_sgr_slc}
  0\ =\ \tilde{\Delta}\sgr{\bsl}
    - \tq^{AB}\twI_B\tD_A\sgr{\bsl}
    - \Ricn{n}{}_{n\bsl}\sgr{\bsl}
    + \lie_{n}\Ricn{n}{}_{\bsl\bsl} \ ,
\end{equation}
where $\tilde{\Delta}$ is the Laplace operator, and $n^{\mu}$ is a null
vector field orthogonal to the slices.  The only part inhomogeneous in
$\sgr{\bsl}$ is a transversal derivative of the Ricci tensor component
$\lie_{n}\Ricn{n}_{\bsl\bsl}|_{\slc}$. If it vanishes on the horizon
\begin{equation}\label{eq:lie-n-Rll}
  \lie_{n}\Ricn{n}_{\bsl\bsl}|_{\ih}\ =\ 0 \ , 
\end{equation}
then according to the vanishing of $\lie_{\bsl}\tq_{AB}$ and
\eqref{eq:lie-sgr} the equation can be rewritten as a PDE
defined on the base space:
\begin{equation}\label{eq:PDE_sgr}
  0\ =\ \hat{\Delta}\sgr{\bsl}
    - \hq^{AB}\hwI_B\hD_A\sgr{\bsl}
    - \Ricn{n}{}_{n\bsl}\sgr{\bsl} \ .
\end{equation}
The equation is now an homogeneous elliptic PDE defined on a compact
manifold. Therefore if $\sgr{\bsl}$ vanishes on some open subset of
$\bas$ then must vanish on the entire horizon. Hence the following is
true:
\begin{cor}\label{cor:part}
  Given an isolated horizon $\ih$ equipped with a symmetry $\bsl$ and
  embedded in a spacetime satisfying the Einstein field
  equations. Assume that
  \begin{equation}\label{c:n_Rll}
    \lie_{n}T_{\bsl\bsl}|_{\slc}\ =\ 0 \ ,
  \end{equation}
  for some null vector field $n$ transversal to the leaves of the
  horizon foliation preserved by the flow $[\bsl]$.
  Then the horizon must belong to one of the following classes:
  \begin{enumerate}[ (i)]
    \item {\it Extremal} isolated horizons: surface gravity vanishes
          everywhere
    \item {\it Non-extremal} ones: $\sgr{\bsl}\neq 0$ on a dense subset
          of $\ih$.
  \end{enumerate}
\end{cor}
The assumed condition is equivalent to \eqref{eq:lie-n-Rll}. When it is
satisfied by the matter fields the partition defined in Corollary
\ref{cor:part} can be used instead of the partition proposed in Section
\ref{sec:ih}. The structure of the constraint (\ref{eq:S_constH})
remains then global.  


\end{document}